\documentclass[aps,prl,preprint,amsfonts,nobibnotes,superscriptaddress]{revtex4}

\usepackage{dcolumn}
\usepackage{amsmath}
\usepackage{amssymb}
\usepackage{graphicx}
\usepackage{bm}
\usepackage{upgreek,xspace}
\usepackage{color}
\usepackage{titlesec}

\begin{document}

\title{Electron-Phonon-Driven Three-Dimensional Metallicity\\ in an Insulating Cuprate}

\vspace{2cm}

\author{Edoardo Baldini}
\affiliation{Institute of Physics, Laboratory for Ultrafast Microscopy and Electron Scattering, \'Ecole Polytechnique F\'ed\'erale de Lausanne, CH-1015 Lausanne, Switzerland}
\affiliation{Institute of Chemical Sciences and Engineering, Laboratory of Ultrafast Spectroscopy, \'Ecole Polytechnique F\'ed\'erale de Lausanne, CH-1015 Lausanne, Switzerland}

\author{Michael A.~Sentef} 
\affiliation{Max Planck Institute for the Structure and Dynamics of Matter, Hamburg, Germany}

\author{Swagata Acharya}
\affiliation{Department of Physics, King's College London, London WC2R 2LS, United Kingdom}

\author{Thomas Brumme}
\affiliation{Max Planck Institute for the Structure and Dynamics of Matter, Hamburg, Germany}
\affiliation{Wilhelm Ostwald Institut of Physical and Theoretical Chemistry, University of Leipzig, Leipzig, Germany}

\author{Evgeniia Sheveleva} 
\affiliation{Department of Physics, University of Fribourg, Chemin du Mus\'ee 3, CH-1700 Fribourg, Switzerland}

\author{Fryderyk Lyzwa}
\affiliation{Department of Physics, University of Fribourg, Chemin du Mus\'ee 3, CH-1700 Fribourg, Switzerland}

\author{Ekaterina Pomjakushina}
\affiliation{Solid State Chemistry Group, Laboratory for Multiscale Materials Experiments, Paul Scherrer Institute, CH-5232 Villigen PSI, Switzerland}

\author{Christian Bernhard} 
\affiliation{Department of Physics, University of Fribourg, Chemin du Mus\'ee 3, CH-1700 Fribourg, Switzerland}

\author{Mark van Schilfgaarde}
\affiliation{Department of Physics, King's College London, London WC2R 2LS, United Kingdom}

\author{Fabrizio Carbone}
\affiliation{Institute of Physics, Laboratory for Ultrafast Microscopy and Electron Scattering, \'Ecole Polytechnique F\'ed\'erale de Lausanne, CH-1015 Lausanne, Switzerland}

\author{Angel Rubio} 
\affiliation{Max Planck Institute for the Structure and Dynamics of Matter, Hamburg, Germany}
\affiliation{Nano-Bio Spectroscopy Group, Departamento de Fisica de Materiales, Universidad del Pa\'is Vasco, 20018 San Sebast\'ian, Spain}
\affiliation{Center for Computational Quantum Physics, The Flatiron Institute, 162 Fifth Avenue, New York, NY 10010, USA}

\author{C{\'e}dric Weber}
\affiliation{Department of Physics, King's College London, London WC2R 2LS, United Kingdom}

\begin{abstract}
\textbf{The role of the crystal lattice for the electronic properties of cuprates and other high-temperature superconductors remains controversial despite decades of theoretical and experimental efforts. While the paradigm of strong electronic correlations suggests a purely electronic mechanism behind the insulator-to-metal transition, recently the mutual enhancement of the electron-electron and the electron-phonon interaction and its relevance to the formation of the ordered phases have also been emphasized. Here, we combine polarization-resolved ultrafast optical spectroscopy and state-of-the-art dynamical mean-field theory to show the importance of the crystal lattice in the breakdown of the correlated insulating state in an archetypal undoped cuprate. We identify signatures of electron-phonon coupling to specific fully-symmetric optical modes during the build-up of a three-dimensional metallic state that follows charge photodoping. Calculations for coherently displaced crystal structures along the relevant phonon coordinates indicate that the insulating state is remarkably unstable toward metallization despite the seemingly large charge-transfer energy scale. This hitherto unobserved insulator-to-metal transition mediated by fully-symmetric lattice modes can find extensive application in a plethora of correlated solids.}
\end{abstract}



\maketitle

The insulator-to-metal transition (IMT) and high-temperature ($T_C$) superconductivity in cuprates are central topics in condensed matter physics \cite{lee2006doping, keimer2015quantum}. A crucial roadblock towards a complete understanding of the IMT and the details of the phase diagram in these compounds lies in the strong-correlation problem. Electron-electron correlations have long been thought to be
the dominant actor responsible for the IMT, whereas the crystal lattice and the electron-phonon coupling have played a secondary role. As a result, much of our present knowledge about the relevant physics of cuprates has been framed around the two-dimensional Hubbard model.

Recently, this purely electronic scenario has been challenged by a body of work. On the theory side, it is believed that the selective modification of bond lengths and angles can trigger a localization-delocalization transition in the undoped parent compounds \cite{cedric_tcmax,gap_trend}, or even lead to a concomitant increase of the superconducting $T_C$ \cite{swagata_prx}. On the experimental side, the interplay between the electron-electron and the electron-phonon interaction has been proposed as an efficient pathway to stabilize superconductivity \cite{mishchenko2004electron,gunnarsson_interplay_2008,gerber2017femtosecond,he2018rapid}. The emergent picture is that electronic correlations and electron-phonon coupling cannot be considered as independent entities in the high-$T_C$ problem, but rather as equally fundamental interactions that can mutually enhance each other.

While this intertwined character of different interactions makes cuprates excellent candidates to benchmark new theories in correlated-electron physics, it also renders these solids a puzzling case to understand \cite{fradkin_colloquium:_2015}. First-principles theoretical descriptions that deal with strong correlations are notoriously difficult to handle, and only now powerful methods are becoming available that render the problem tractable on modern computers \cite{swagata_prx, golevz2019dynamics}. At the same time, experimental progress in disentangling intricate interactions relies on the development of novel spectroscopic techniques. In particular, driving complex systems out-of-equilibrium and monitoring their real-time behavior with ultrafast probes \cite{giannetti_ultrafast_2016} has evolved as a promising strategy to uncover the relevance of various microscopic degrees of freedom and the mutual forces between them \cite{sentef_examining_2013}.

The application of ultrafast methods to undoped cuprates has revealed preliminary details on the dynamics underlying the IMT. This was accomplished by photodoping particle-hole pairs in the CuO$_2$ planes with a short laser pulse while monitoring the change in the optical absorption spectrum with a delayed continuum probe \cite{ref:okamoto_1, ref:okamoto_2}. The extremely fast timescale (40-150 fs) associated with the rise of the low-energy Drude response was found to be imprinted onto the dynamical evolution of the optical charge-transfer (CT) excitation in the visible range. Within 200 fs from its formation, the mobile charges freeze into localized mid-gap states owing to the concomitant action of polar lattice modes and spin fluctuations \cite{mishchenko2008charge}. Finally, the self-trapped carriers release energy in the form of heat over a picosecond timescale.

Despite their pioneering contribution, these works have left several fundamental questions unanswered. First, the use of cuprate thin films has hindered the study of the charge dynamics along the crystallographic $c$-axis. Hence, it is unknown whether the transient metallic state has a purely two-dimensional nature or whether it also involves a certain degree of interlayer transport. Furthermore, the high temperature employed in these experiments has masked the observation of possible bosonic collective modes that cooperate with the charge carriers to induce the IMT. 

Here we combine ultrafast optical spectroscopy and first-principles calculations to unravel the intricate role of the electron-phonon coupling in the stability of the insulating state of a prototypical cuprate parent compound. By measuring the non-equilibrium response of different elements of the optical conductivity tensor, we reveal that rapid injection of particle-hole pairs in the CuO$_2$ planes leads to the creation of a three-dimensional (3D) metallic state that has no counterpart among the chemically doped compounds, and is accompanied by a complex motion of the ionic positions along the coordinates of fully-symmetric modes. The information gleaned from our experiment about the phonons that strongly couple to the mobile charges is supported by a state-of-the-art theoretical framework that unveils a striking instability of the insulating state against the displacement of the same lattice modes. These findings indicate that the light-induced IMT in cuprates cannot be interpreted as a purely electronic effect, calling for the involvement of intertwined degrees of freedom in its dynamics. More generally, these results open an avenue toward the phonon-driven control of the IMT in a wide class of insulators in which correlated electrons are strongly coupled to fully-symmetric lattice modes.

\section{Results}

\subsection*{Crystal Structure and Equilibrium Optical Properties}

As a model material system we study La$_2$CuO$_4$ (LCO), one of the simplest insulating cuprates exhibiting metallicity upon hole doping. In this solid, the two-dimensional network of corner-sharing CuO$_4$ units is accompanied by two apical O atoms below and above each CuO$_4$ plaquette. As a result, the main building blocks of LCO are CuO$_6$ octahedra (Fig.~1 $A$) that are elongated along the $c$-axis due to the Jahn-Teller distortion. The unit cell of LCO is tetragonal above and orthorhombic below 560 K. A simplified scheme of the electronic density of states is shown in Fig.~1 $B$ (left panel). An energy gap ($\Delta_{CT}\sim$2 eV) opens between the filled O-2$p$ band and the unoccupied Cu-3$d$ upper Hubbard band (UHB), thus being of the CT type. In contrast, the occupied Cu-3$d$ lower Hubbard band (LHB) lies at lower energy.


First, we present the optical properties of LCO in equilibrium. Figure 1 $C$ shows the absorptive part of the optical conductivity ($\sigma_1$), measured via ellipsometry. The in-plane response ($\sigma_{1a}$, solid violet curve) is dominated by the optical CT gap at 2.20 eV \cite{ref:uchida_LSCO, ref:falck}. This transition is a non-local resonant exciton that extends at least over two CuO$_4$ units. The strong coupling to the lattice degrees of freedom causes its broadened shape \cite{ref:falck, shen_missing_2004, ref:ellis}. As such, this optical feature can be modeled as involving the formation of an electron-polaron and a hole-polaron, coupled to each other by a short-range interaction \cite{ref:falck, mann2015probing}. At higher energy (2.50$-$3.50 eV), the in-plane spectrum results from charge excitations that couple the O-2\textit{p} states to both the Cu-3\textit{d} states in the UHB and the La-5\textit{d}/4\textit{f} states. In contrast, the out-of-plane optical conductivity ($\sigma_{1c}$) is rather featureless and its monotonic increase with energy is representative of a particle-hole continuum. This spectral dependence reflects the more insulating nature of LCO along the \textit{c}-axis, which stems from the large interlayer distance between neighboring CuO$_2$ planes. As a consequence, over an energy scale of 3.50 eV, charge excitations in equilibrium are mainly confined within each CuO$_2$ plane.

\subsection*{Photoinduced Three-Dimensional Metallic State}

We now reveal how these optical properties of LCO modify upon above-gap photoexcitation. To this aim, we tune the photon energy of an intense ultrashort laser pulse above the in-plane optical CT gap energy (blue arrow in the left panel of Fig.~1 $B$), photodoping particle-hole pairs into the CuO$_2$ planes. We explore an excitation regime between 0.023 and 0.075 photons/Cu atom in order to exceed the threshold density needed in LCO for the formation of in-plane metallic conductivity \cite{ref:okamoto_2}. We then use a continuum probe to map the pump-induced changes of the optical response over the CT energy scale (schematic in the right panel of Fig. 1 $B$ and shaded area in Fig. 1 $C$). Unlike previous experiments \cite{ref:okamoto_1, ref:okamoto_2, novelli2014witnessing}, the combination of an (100)-oriented single crystal, an accurate polarization-resolved pump-probe analysis, and low temperature, allows us to identify hitherto undetected details of the light-induced IMT. 

Figures 2 $A,B$ show the spectro-temporal evolution of the $a$- ($\Delta\sigma_{1a}$) and $c$-axis ($\Delta\sigma_{1c}$) differential optical conductivity in response to in-plane photoexcitation. Transient spectra at representative time delays are displayed in Fig. 2 $C,D$. These data are obtained from the measured transient reflectivity through a differential Lorentz analysis \cite{novelli2012ultrafast, borroni2017coherent}, which avoids the systematic errors of Kramers-Kronig transformations on a finite energy range. 

Injecting particle-hole pairs in the CuO$_2$ planes produces a sudden reduction in $\Delta\sigma_{1a}$ close to the optical CT excitation and to its delayed increase to positive values in the 1.80-2.00 eV range (Fig. 2 $A,C$). As explained in previous studies \cite{ref:okamoto_1, ref:okamoto_2}, this behavior stems from the pump-induced redistribution of spectral weight from high to low energy due to several processes, among which the ultrafast emergence of in-plane metallicity, charge localization in mid-gap states, and lattice heating. In particular, the latter causes the first derivative-like shape that gradually arises after several hundreds of femtoseconds and becomes dominant on the picosecond timescale (compare the curve at 1.50 ps in Fig. 2 $B$ and $\Delta\sigma_{1a}$ of Fig. S2 $C$ produced by the lattice temperature increase).

The same photodoping process also modifies $\Delta\sigma_{1c}$ (Fig. 2 $B,D$), which is unveiled here for the first time. At 0.10 ps, a crossover between a reduced and an increased $\Delta\sigma_{1c}$ emerges around 2.00 eV. Subsequently, the intensity weakly drops over the whole spectrum and relaxes into a negative plateau that persists for picoseconds while the system thermalizes to equilibrium. The response is featureless and one order of magnitude smaller than its in-plane counterpart. Here we show that this suppressed background is key to unraveling invaluable information of the intricate dynamics of LCO.

First, we compare the temporal evolution of $\Delta\sigma_1$ along the two crystallographic axes and focus on the dynamics close to zero time delay. Figure 2 $E$ displays a representative trace of $\Delta\sigma_{1a}$ obtained through integration around the optical CT feature. The intensity drops within $\sim$0.17 ps, a timescale that is significantly longer than our resolution ($\sim$0.05 ps). The subsequent relaxation spans tens of picoseconds. A close inspection around zero time delay (inset of Fig. 2 $E$) reveals a resolution-limited signal that emerges only partially before being buried under the pronounced intensity drop. Figure 2 $F$ shows the out-of-plane response (integrated over the low-energy region of Fig. 2 $D$), which shares important similarities with the in-plane signal: The intensity suppression is also complete within $\sim$0.17 ps and comprises a resolution-limited feature that perfectly mirrors the one along the $a$-axis. This signal cannot originate from a leakage of the other probe polarization channel, as the shape of the $\Delta\sigma_{1c}$ spectrum has no fingerprint of the in-plane CT exciton. Furthermore, since the resolution-limited temporal response is only observed in LCO and over a broad spectral range away from the pump photon energy, it cannot be a pump-induced artifact (see $\S$S4). Conversely, the combination of high time resolution and a continuum probe allows us to ascribe this feature to the signature imprinted onto the optical CT energy scale by a light-induced metallic state.


This conclusion naturally emerges through direct inspection of our spectro-temporal response. Previous pump-probe measurements covering the 0.10-2.20 eV range identified an ultrafast transfer of spectral weight from the above-gap to the below-gap region, with the establishment of Drude conductivity within the CuO$_2$ planes \cite{ref:okamoto_2}. Due to this spectral weight transfer, the high-energy region of the spectrum becomes sensitive to the build-up and relaxation dynamics of the itinerant carrier density. The transient metallic behavior manifests itself with a pulsed signal that modifies the optical CT gap feature and decays within 150 fs from the arrival of the pump pulse. The sign and shape of the transient metallic response depend on the nature of the optical nonlinearities induced by the delocalized carriers on the high-energy scale. In this respect, the sharp feature that we observe in our temporal traces of Fig. 2 $E, F$ is in excellent agreement with the evolution of the in-plane Drude conductivity found in these previous experiments. More importantly, the rise of $\Delta\sigma_{1c}$ in our data closely mimics that in $\Delta\sigma_{1a}$, indicating that the metallic state has an unexpected 3D character. Quantitative information is obtained through a systematic global fit analysis of the temporal dynamics along both crystallographic axes. An accurate fit is accomplished only through a model based on that proposed in Ref. \cite{ref:okamoto_2}. A Gaussian function representing the metallic state captures the fast-varying signal during the rise of the response, whereas a subsequent multi-exponential relaxation comprises contributions from charge localization in mid-gap states and lattice heating effects. Details are given in $\S$S6; here we only present fits to the traces of Fig. 2 $E,F$, which are overlapped as dashed black lines. 

Besides leaving a characteristic signature in the time domain, the 3D metallic state also influences the ultrafast spectral response of LCO. A similar behavior appears in both the absorptive and dispersive components of $\Delta\sigma_{a}$ and $\Delta\sigma_{c}$ (Figs. S10-S11), but with a time lag between the two directions. This suggests that the optical nonlinearities induced by the itinerant carriers onto the high-energy optical response of LCO follow distinct dynamics along the $a$- and the $c$-axis.

We stress that this 3D metallic state in photodoped LCO is significantly different from the case of chemically-doped LCO \cite{eckstein2013photoinduced}. In the latter, 3D metallicity is suppressed up to doping levels as high as $p$ = 0.12 (i.e. well above our photodoping density), and only in overdoped samples a well-defined Drude response emerges \cite{ref:uchida_caxis}. In contrast, our findings break the scenario of an ultrafast IMT solely governed by two-dimensional quasiparticles in the CuO$_2$ planes.

\subsection*{Dynamics of the Collective Modes}

As a next step, we search for possible collective modes that strongly couple to the mobile carriers and clarify their involvement in the IMT. In this respect, we note that the lack of a significant background in $\Delta\sigma_{1c}$ uncovers a pronounced oscillatory pattern that emerges from the rise of the response and persists during its decay (inset of Fig. 2 $F$). This coherent beating is due to collective modes displaced through the delocalized carrier density \cite{li2013optical, mann2016probing}. Similar oscillations also appear in the $a$-axis polarization channel (Fig. 2 $E$), but the contrast to resolve them is lower owing to the huge relaxation background.




We assign the modes coupled to the in-plane charge density by applying a Fourier transform analysis to the original background-free transient reflectivity data to maintain high accuracy. The results, shown in Fig. 3 $A,B$, reveal that five bosonic excitations (labelled as $A_g(1-5)$) participate in the non-equilibrium response. Their energies match those of the five $\mathrm{A_g}$ phonons reported in orthorhombic LCO by spontaneous Raman scattering \cite{ref:nimori}. We characterize their eigenvectors by calculating the phonon spectrum of LCO via density-functional theory (see Methods section). Figure~3 $C$ shows the involved ionic displacements in the first half cycle of the different $\mathrm{A_g}$ phonons. Modes $A_g(1)$ and $A_g(2)$ exhibit staggered rotations of CuO$_6$ octahedra. In particular, $A_g(1)$ is the soft phonon of the orthorhombic-to-tetragonal transition. $A_g(3)$ and $A_g(4)$ present large $c$-axis displacements of the La atom, which in turn modify the La-apical O distance. The only difference between them lies in the displacement of the apical O: while its out-of-plane motion is the same, its in-plane motion occurs in opposite direction. Lastly, $A_g(5)$ is the breathing mode of the apical O. The Fourier transform indicates that all phonons modulate the out-of-plane response (Fig.~3 $B$, brown curve), whereas only $A_g(1)$, $A_g(3)$ and $A_g(4)$ are unambiguously resolved in the in-plane signal (Fig.~3 $B$, violet curve). Furthermore, since $A_g(1)$, $A_g(2)$ and $A_g(4)$ are characteristic phonons of orthorhombic LCO \cite{ref:nimori}, their presence denotes that the trajectory followed by the lattice after photodoping does not evolve through the structural phase transition. Finally, no modes with symmetry other than $A_g$ appear in our data. While this result is natural when the probe is $c$-axis-polarized because of symmetry arguments, more noteworthy is the in-plane probe polarization case. According to Raman selection rules, modes of $B_{1g}$ symmetry should also emerge in this polarization configuration \cite{mansart2013coupling}. Their absence can be attributed either to the short lifetime of the $B_{1g}$ component of the real charge-density fluctuation driving the coherent lattice response \cite{li2013optical}, or to a weak Raman cross-section in the probed spectrum.



To establish which $A_g$ modes preferentially couple to the in-plane photodoped carriers, we also excite LCO with a light field polarized along the \emph{c}-axis. Despite keeping the carrier excitation density constant, this pump scheme causes a smaller drop in the transient signal amplitude and a weaker modulation depth due to the coherent lattice modes (Fig. S7). Fourier transforming the background-free data (Fig. 3 $A,B$, green curves) establishes that only $A_g(1)$ and $A_g(2)$ are efficiently triggered by the out-of-plane electronic density, whereas $A_g(3)$ and $A_g(4)$ are strongly suppressed compared to the in-plane photoexcitation scheme.
This suggests that periodic structural elongations and compressions of the CuO$_6$ octahedra along the \textit{c}-axis through the La atom are favorably triggered by photodoping charges within the CuO$_2$ planes. This aspect can be explained by noting that excitation across the optical CT gap promotes electrons in the UHB and holes in the O-2$p$ band, thus locally removing the Jahn-Teller distortion on the CuO$_6$ octahedra. In the past, this picture has been explored theoretically for chemical (hole) doping, showing how the apical O approaches the Cu$^{2+}$ ions to gain attractive electrostatic energy. Consistent with this idea, the dominant mode in our experiment involves coherent displacements of the apical O and the La atoms along the $c$-axis, i.e. an oscillating motion that likely follows the destabilization of the Jahn-Teller distortion. 

\subsection*{Role of the Electron-Phonon Coupling in the Insulator-to-Metal Transition}

Our results indicate that the ultrafast 3D metallization of LCO is accompanied by a complex structural motion that is strongly coupled to the delocalized carriers. This motivates us to study theoretically whether the ionic displacements along the relevant lattice mode coordinates can also influence the electronic properties of LCO. To this end, we perform advanced calculations using dynamical mean-field theory on top of a quasiparticle self-consistent $GW$ approach (QS\emph{GW}+DMFT, see Methods). This recently-developed method offers a non-perturbative treatment of the local spin fluctuations that are key for the electronic properties of undoped cuprates. We benchmark this technique on the in-plane equilibrium optical properties of LCO. Figure 1 $C$ shows the calculated $\sigma_{1a}$ for an undisplaced unit cell (dashed violet curve). The shape of the calculated optical spectra have a remarkable quantitative agreement with the experimental data, strongly validating our theory. Our approach accurately captures the $d$-$p$ correlations that are pivotal for the Madelung energy \cite{non_local}, refining the description of the optical absorption spectrum of correlated insulators \cite{comanac2008optical}.


Using this method, we address the influence of distinct lattice modes on the electronic structure and optical properties of LCO. We compute the single-particle spectral properties and optical conductivity within the frozen-phonon approximation, i.e. by statically displacing the ions in the unit cell along the coordinates of relevant Raman-active modes. While this adiabatic method can only provide information on the electron-phonon coupling in the electronic ground state, it represents a first important step to elucidate how specific atomic motions affect the electronic properties of this correlated insulator. Figure 4 $A-C$ shows some representative results, whereas Figs. S13-S16 provide the full analysis. To emphasize the impact of the different ionic motions on the optical conductivity, we show spectra obtained upon displacing the unit cell of LCO by 0.04 $\AA$ along different phonon coordinates. However, smaller values of frozen lattice displacement yield similar results. Surprisingly, we observe that displacements along each of the $A_{g}$ modes induce net metallization along the $a$-axis (Fig. 4 and Figs. S13-S14). The system evolves into a bad metal with an incoherent quasiparticle peak in the single-particle spectral function and a broad Drude response in $\sigma_{1a}$. In contrast, modes with symmetries other than $A_{g}$ cause no metallic instability within the CuO$_2$ planes. We also extend these calculations to the $c$-axis optical response and test how the out-of-plane insulating state of LCO reacts against the same lattice displacements. Capturing the correct onset of the equilibrium optical conductivity along the $c$-axis is a very demanding computational task, as the use of electron-hole screening vertex corrections becomes crucial in the presence of very small bandwidths \cite{go2015spatial}. Our current QS$GW$+DMFT theory level does not incorporate such vertex corrections and the resulting $\sigma_{1c}$ is blueshifted compared to the experimental spectrum (Fig. S15). Nevertheless, our approach is sufficiently robust to elucidate the effect of different ionic motions on the $c$-axis insulating state. The results, shown in Fig. S16 as violet curves, confirm the trends reported for the $A_{g}$ modes, establishing that displacements along their eigenvectors cause net metallization also along the $c$-axis. These findings have three important implications: i) The spectral region covered by our experiment is sensitive to the displacements of modes with any symmetry, ruling out the possibility of a weak Raman cross-section behind the lack of $B_{1g}$ modes in our experiment; ii) Displacements along the coordinates of modes with a fully-symmetric representation can induce an IMT; iii) The lattice-driven metallic state has a 3D nature.



Let us discuss the possible contributions that can explain the observed metallicity in the $A_{g}$-displaced structures: effective doping, change of screening, and modification of orbital overlaps. Regarding the doping into Cu-3$d$ states, displacements along the coordinates of all Raman-active modes (except B$_{3g}$) lead to an increase in the hole density in the $d$ states relative to the undisplaced case (Table S2 of the Supplemental Materials). In the case of the $B_g$ modes, this effective doping yields $\sim$0.4$\%$ holes for B$_{1g}$, $\sim$0.3$\%$ holes for B$_{2g}$, and $\sim$1.5$\%$ electrons for B$_{3g}$. In the case of the $A_g$ modes (except $A_g(1)$), the hole doping reaches values larger than $\sim$3$\%$. It appears that all the Raman-active modes that cause significant effective doping in Cu-3$d$ states transfer spectral weight to lower energies, leading to weak metallization. Enhanced screening in the displaced structures could also lead to the breakdown of the insulating state via modified effective Hubbard $U$ and CT energies. However, our estimates of the CT energy for the displaced structures show very small variation compared to that of undisplaced LCO (Table S1 of the Supplemental Materials). While these changes are typically larger for modes that lead to metallization, the rather small modification of the CT energy alone is unlikely to cause the loss of the insulating state. Finally, by the principle of exclusion, the orbital overlaps and thus the hopping integrals between Cu-3$d$ and O-2$p$ orbitals remain as the main explanation for the emerging metallic state, due to their high sensitivity to certain lattice displacements. While we do not explicitly compute values for renormalized hoppings here, the momentum-resolved spectral functions for the displaced structures (Fig. S13) strongly suggest that modified hoppings are indeed the most important ingredient in the observed metallization.

%
%

\section{Discussion}

The fundamental and technological implications of our results are noteworthy. On the theory side, we have reported a hitherto undetected type of IMT, which applies to pure Mott/CT insulators (i.e. devoid of coexisting charge-orbital orders \cite{perfetti2006time, hellmann2012time, de2013speed} and not lying in proximity to a structural transformation \cite{cavalleri2004evidence, kubler2007coherent, morrison2014photoinduced}). As such, this IMT can be extended to a wide class of correlated solids (e.g., NiO, iridates, etc.). Moreover, in the specific case of cuprates, our data enrich the debate around the role and structure of the electron-phonon interaction \cite{lanzara2001evidence, ronning2005anomalous, graf2007universal, gedik2007nonequilibrium, gunnarsson_interplay_2008, moritz2009effect, johnston, rameau, he2018rapid}. In particular, an old puzzle in the field regards the evolution of the quasiparticle-like excitations from the undoped Mott insulator to the doped compounds \cite{shen_missing_2004,mannella_nodal_2005}. This problem is closely related to the correct identification of the chemical potential and its behavior upon hole- or electron-doping. A possible solution of this paradox was proposed by noting that the coherent quasiparticle scenario fails to describe the Mott insulating state and that a Franck-Condon-type of broadening contributes to the lineshape of the main band in the single-particle excitation spectrum \cite{shen_missing_2004}. In the Franck-Condon scenario, the true quasiparticle peak at half filling has a vanishingly small weight and the spectrum is dominated by incoherent sidebands due to shake-off excitations stemming from the coupling between the electrons and bosonic collective modes.

Our results add new insights to this long-standing problem. First, the combination of our equilibrium optical data and calculations reinforces the idea that polar lattice modes cooperate with the electronic correlations to freeze quasiparticles and stabilize the insulating state of LCO \cite{ref:rosch, mishchenko2008charge, ref:okamoto_1, ref:okamoto_2, novelli2014witnessing}. This is evidenced by the tail of optical spectral weight extending down to 1.00 eV in Fig. 1 $C$, at odds with the 1.80 eV electronic-only gap retrieved by our simulations. This discrepancy can be explained by noting that our theory does not account for either the electron-phonon or the electron-hole interactions. As such, the tail emerging below 1.80 eV in $\sigma_{1a}$ is most likely due to the interplay between excitonic states and polar lattice modes \cite{toyozawa2003optical}. This observation would clarify why the energy gap (and the barrier to metallicity) relevant to the transport and thermodynamic properties of insulating cuprates is much smaller that the optical CT gap (for LCO the transport gap is estimated around 0.89 eV, at the onset of the tail in $\sigma_{1a}$) \cite{ono2007strong, xiang2009intrinsic, moskvin2011true}. The same polar lattice modes would also be responsible for the charge-phonon coupling revealed by the previous photodoping experiments on LCO \cite{ref:okamoto_1, ref:okamoto_2}, shaping the structure of the mid-infrared absorption bands \cite{mishchenko2008charge}. On top of this, our study uncovers that (real) Raman-active lattice displacements can instead induce quasiparticle delocalization and trigger an IMT. Taken together, these results constitute additional proof for the role of the crystal lattice in both quasiparticle dressing and stability toward metallization, indicating that the strong-correlation problem is incomplete without phonons.

On the applied side, the important upshot from our findings is that excitation of specific Raman-active modes bears huge potential for the control of the IMT in correlated insulators. In recent years, the notion of nonlinear phononics has opened an avenue toward the lattice-mediated control of electronic properties in a wide variety of doped correlated materials \cite{Mankowsky2016, subedi}. The underlying mechanism involves the resonant excitation of large-amplitude infrared-active modes, whose oscillation displaces the crystal along the coordinates of coupled Raman-active modes. Extending these studies to the insulating parent compounds by employing narrow-band THz fields and sum-frequency ionic Raman scattering \cite{maehrlein_terahertz_2017, terashige2019doublon} will realize the mode-selective control of the IMT in the electronic ground state, paving the way to the use of correlated insulators in fast room-temperature devices. Finally, our joint experimental-theoretical effort points to rational-design strategies for novel quantum materials that exhibit a subtle interplay of electron-electron and electron-phonon interactions.

\begin{figure*}[t]
	\centering
	\includegraphics[width=\columnwidth]{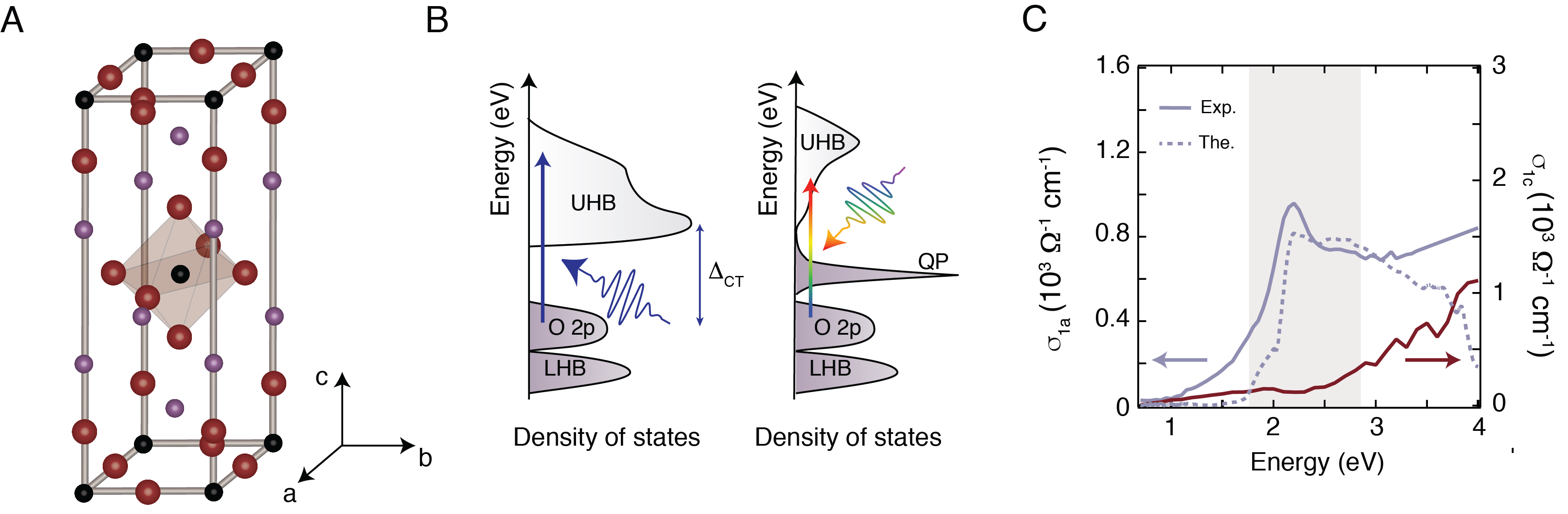}
	\caption{(\textit{A}) Crystallographic structure of La$_2$CuO$_4$ in its low-temperature orthorhombic unit cell. The Cu atoms are depicted in black, the O atoms in red, and the La atoms in violet. The brown shading emphasized the CuO$_6$ octahedron in the center. (\textit{B}) Schematic representation of the interacting density of states in undoped insulating (left panel) and photodoped metallic (right panel) La$_2$CuO$_4$. The O-2$p$, lower Hubbard band (LHB), upper Hubbard band (UHB), and quasiparticle (QP) peak are indicated. In the insulating case, the optical charge-transfer gap ($\Delta_{\text{CT}}$) is also specified. The blue arrow indicates the 3.10 eV pump pulse, which photodopes the material and creates particle-hole pairs across the charge-transfer gap. The multicolored arrow is the broadband probe pulse, which monitors the high-energy response of the material after photoexcitation. (\textit{C}) Real part of the optical conductivity at 10 K, measured with the electric field polarized along the $a$- (violet solid curve) and the $c$-axis (brown solid curve). The shaded area represents the spectral region monitored by the broadband probe pulse in the nonequilibrium experiment. The theory data for the in-plane response are shown as a violet dashed curve. The $a$-axis response comprises a well-defined peak in correspondence to the optical charge-transfer gap around 2.20 eV and a tail extending toward low energies down to 1.00 eV. In contrast, the $c$-axis response is featureless and increases monotonically with increasing energy, as expected from a particle-hole continuum.}\label{fig:Figure1}
\end{figure*}
\clearpage
\newpage

\begin{figure*}[t]
	\centering
	\includegraphics[width=\columnwidth]{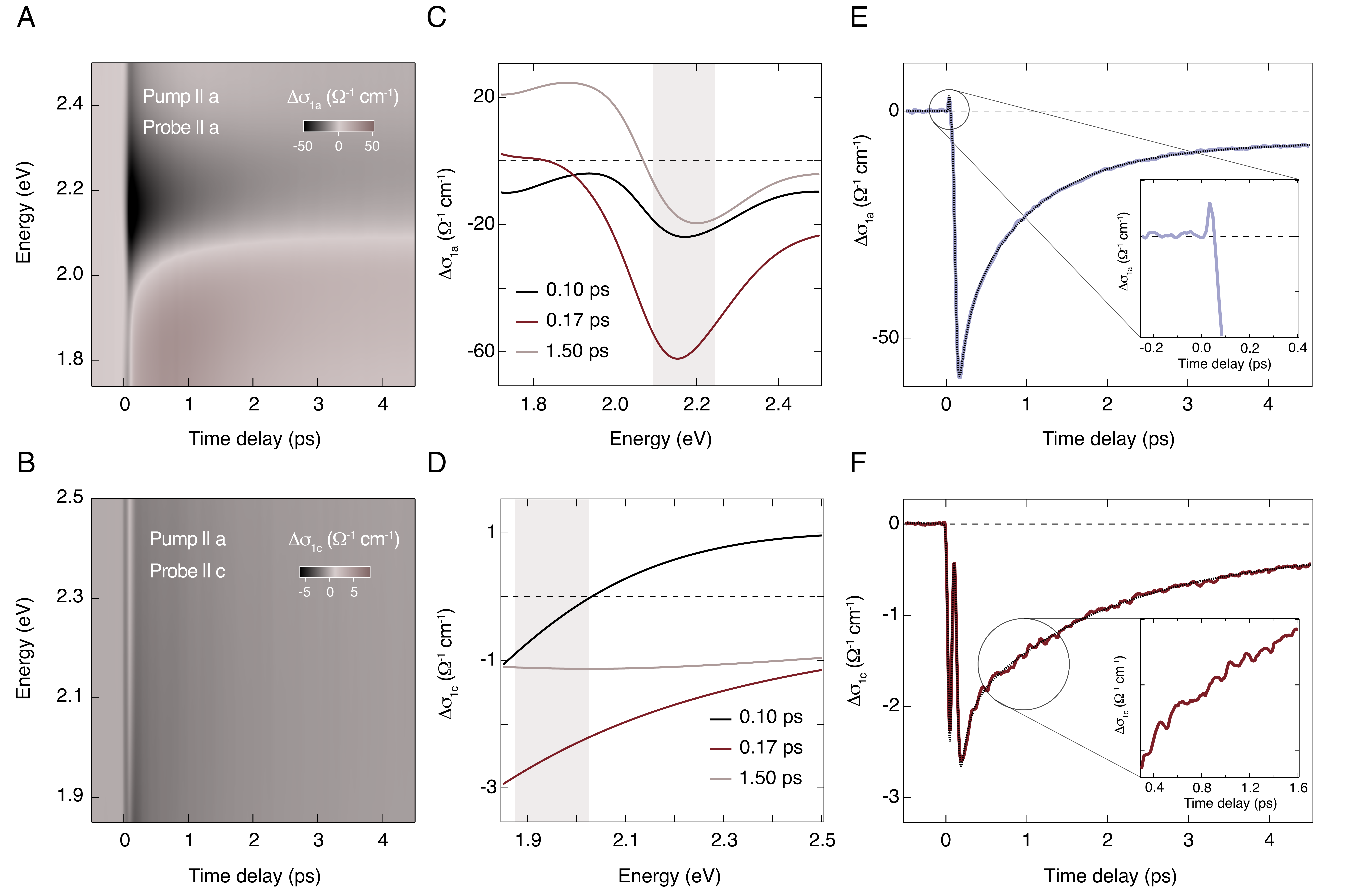}
	\caption{(\textit{A,B}) Color-coded maps of the differential optical conductivity ($\Delta\sigma_{1}$) at 10 K with in-plane pump polarization and ($A$) in-plane, ($B$) out-of-plane probe polarization, as a function of probe photon energy and pump-probe time delay. The pump photon energy is 3.10 eV and the excitation photon density is $x_{ph}$ $\sim$ 0.06 photons per copper atom. For in-plane probe polarization ($A$), we observe a significantly reduced $\Delta\sigma_{1}$ above the optical CT edge at 1.80 eV, due to spectral weight redistribution to lower energies. For out-of-plane probe polarization ($B$), the depletion in $\Delta\sigma_{1}$ is considerably weaker and rather featureless. Oscillatory behavior is visible in the color-coded map hinting at coherently-excited phonon modes. (\textit{C,D}) Snapshots of the same data as in $A,B$ at three different pump-probe time delays during the rise (0.10 ps and 0.17 ps) and the relaxation (1.50 ps) of the response. (\textit{E,F}) Temporal traces of $\Delta\sigma_{1}$ along the $a$- and $c$-axis. Each temporal trace results from the integration over the energy window indicated by the shaded areas in $C,D$. The in-plane trace ($E$) shows a dramatic suppression and slow recovery without clearly visible coherent oscillations. A small peak emerges in the rise of the response, due to the light-induced metallic state (inset). The out-of-plane trace ($F$) shows a clear signal that mimics the fast in-plane response, but relaxes with a tail exhibiting pronounced coherent oscillations (highlighted in the inset).}\label{fig:side}
\end{figure*}
\clearpage
\newpage

\begin{figure*}[t]
	\centering
	\includegraphics[width=0.8\columnwidth]{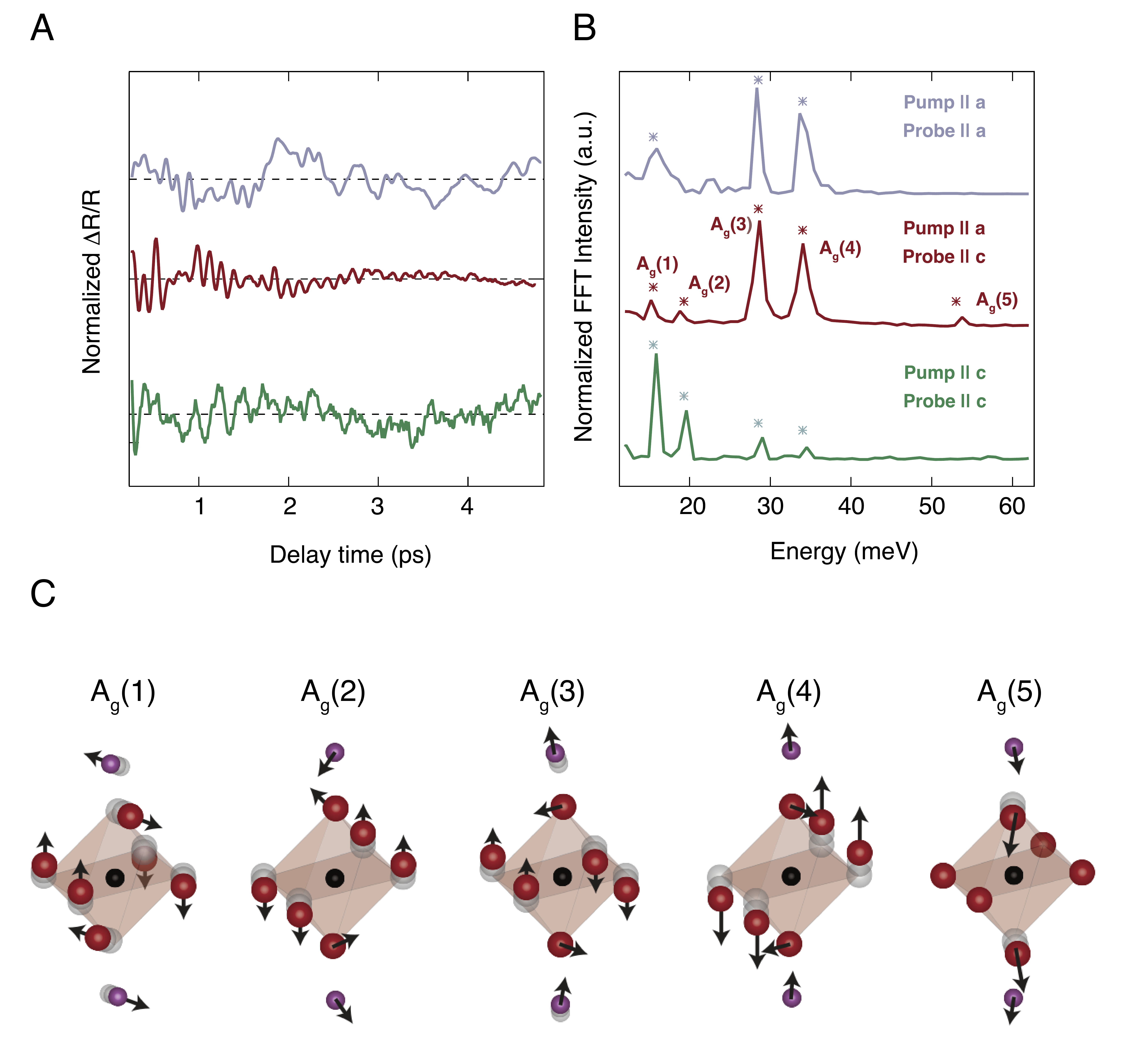}
	\caption{(\textit{A}) Residual reflectivity change (normalized to the largest amplitude) after subtraction of the recovering background, exhibiting coherent oscillations due to collective bosonic modes. (\textit{B}) Fast Fourier transform of data in $A$. The data in panels $A$ and $B$ refer to different pump and probe polarizations as indicated in $B$. The traces have been selected in the probe spectral region that maximizes the oscillatory response (2.00-2.20 eV for the violet curve, 1.80-2.20 eV for the brown and the green curves). Different polarizations show the presence of a set of totally symmetric ($A_g$) phonon modes of the orthorhombic crystal structure. The asterisks in panel $B$ indicate the phonon energy as measured by spontaneous Raman scattering \cite{ref:nimori}. (\textit{C}) Calculated eigenvectors of the five modes of $A_g$ symmetry. Black atoms refer to Cu, red atoms to O and violet atoms to La. Modes $A_g(1)$ and $A_g(2)$ involve staggered rotations of CuO$_6$ octahedra. Modes $A_g(3)$ and $A_g(4)$ present large $c$-axis displacements of the La atom, which in turn modify the La-apical O distance. The only difference between them lies in the displacement of the apical O: while its out-of-plane motion is the same, its in-plane motion occurs in opposite direction. Mode $A_g(5)$ is the breathing mode of the apical O. The phonon spectrum has been computed using density-functional theory.}\label{fig:Figure3}
\end{figure*}
\thispagestyle{empty} 
\clearpage
\newpage

\begin{figure*}
	\centering
	\includegraphics[width=\columnwidth]{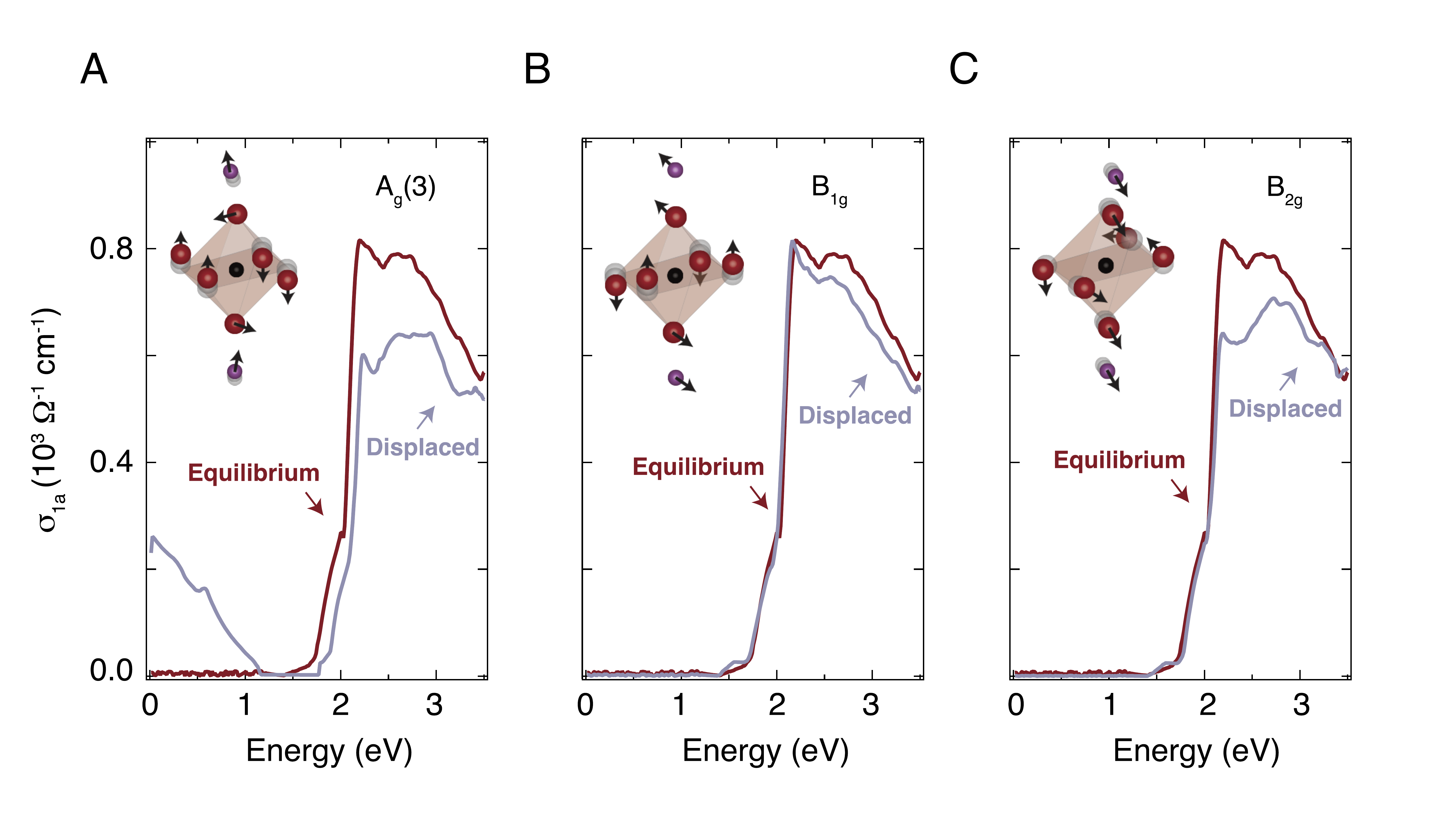}
	\caption{($A-C$) Many-body calculations of the in-plane optical conductivity for the La$_2$CuO$_4$ unit cell. Comparison between the response for the undisplaced structure (brown curve) and the response for the structure displaced by 0.04 $\AA$ along the phonon coordinates indicated in each panel (violet curves). For displacements along totally-symmetric modes (an example is shown in $A$), a metallic state emerges and gives rise to Drude spectral weight below $\sim$ 1.00 eV. In contrast, for displacements along $B_g$ modes (examples are given in $B,C$), there is no metallization and hence no impact on the low-energy spectral weight inside the optical charge-transfer gap.}
	\label{fig:Figure4}
\end{figure*}
\clearpage
\newpage

\section{Methods}

\subsection*{Single crystal growth and characterization}
	
	Polycrystalline LCO was prepared by a solid state reaction. The starting materials La$_2$O$_3$ and CuO with  99.99\% purity were mixed and ground. This process was followed by a heat treatment in air at 900$^\circ$C-1050$^\circ$C for at least 70 hours with several intermediate  grindings. The phase purity of the resulting compound was checked  with  a  conventional x-ray diffractometer. The resulting powder was hydrostatically pressed into rods (7 mm in diameter) and subsequently sintered at 1150$^\circ$C for 20 hours. The crystal growth was carried out using an optical floating zone furnace (FZ-T-10000-H-IV-VP-PC, Crystal System Corp., Japan) with four 300 W halogen lamps as heat sources. The growing conditions were as follows: the growth rate was 1 mm/h, the feeding and seeding rods were rotated at about 15 rpm in opposite directions to ensure the liquid's homogeneity, and an oxygen and argon mixture at 3 bar pressure was applied during the growth. The as-grown crystals were post-annealed at 850$^\circ$C in order to release the internal stress and to adjust the oxygen content. One crystal was oriented in a Laue diffractometer, cut along a plane containing the \textit{a} and \textit{c} axes and polished to optical quality. Initially, the N\'eel temperature was determined to be $\mathrm{T_N}$ = 260 K, which corresponds to a doping $\delta$ = 3 $\times$ 10$^{-3}$ and a hole content $p$ = 6 $\times$ 10$^{-3}$. For this reason, the crystal was annealed for 48 hours to remove part of the excess oxygen. After the treatment, $\mathrm{T_N}$ increased to 307 K, which well agrees with the typical value found in purely undoped compounds.
	
	\subsection*{Ellipsometry}
	We used spectroscopic ellipsometry to measure the complex dielectric function of the sample, covering the spectral range from 0.80 eV to 6.00 eV. The experiments were performed using a Woollam VASE ellipsometer. The LCO single crystal was mounted in a helium flow cryostat, allowing measurements from room temperature down to 10 K. The measurements were performed at $<$10$^{-8}$ mbar to prevent measurable ice condensation onto the sample. Anisotropy corrections were performed using standard numerical procedures.
	
\subsection*{Ultrafast broadband optical spectroscopy}
	
	For the ultrafast optical experiments, we used an amplified laser system operating at a repetition rate of 3 kHz \cite{baldini2016versatile, baldini2017clocking, baldini2018lattice}. The set-up was based on a Ti:Sapphire oscillator, pumped by a continuous-wave Nd:YVO$_4$ laser, emitted sub-50 fs pulses at 1.55 eV with a repetition rate of 80 MHz. The output of the oscillator seeded a cryo-cooled Ti:Sapphire amplifier, which was pumped by a Q-switched Nd:YAG laser. This laser system provided $\sim$45 fs pulses at 1.55 eV. One third of the output, representing the probe beam, was sent to a motorized delay line to set a controlled delay between pump and probe. The 1.55 eV beam was focused on a 3 mm-thick CaF$_2$ cell using a combination of a lens with short focal distance and an iris to limit the numerical aperture of the incoming beam. The generated continuum covered the 1.77-2.90 eV spectral range. The probe was subsequently collimated and focused onto the sample through a pair of parabolic mirrors at a small angle from normal incidence. The remaining two thirds of the amplifier output, representing the pump beam, were frequency doubled to 3.10 eV in a $\beta$-barium borate crystal and directed toward the sample under normal incidence. Along the pump path, a chopper with a 60 slot plate was inserted, operating at 1.5 kHz and phase-locked to the laser system. Both pump and probe were focused onto the sample on spots of dimensions \mbox{120 $\mathrm{\mu m}$ $ \times$ 87 $\mathrm{\mu m}$} for the pump and 23 $\mathrm{\mu m}$ $\times$ 23 $\mathrm{\mu m}$ for the probe. The sample was mounted inside a closed-cycle cryostat, which provided a temperature-controlled environment in the range 10-340 K. The reflected probe was dispersed by a fiber-coupled 0.3 m spectrograph and detected on a shot-to-shot basis with a complementary metal-oxide-semiconductor linear array. Before the data analysis, the transient reflectivity matrix was corrected for the group velocity dispersion of the probe. It is important to note that the probe beam dispersion was not a limiting factor for the time resolution of the setup, since it was given on the detection side by the much smaller effective pulse duration per detector pixel. As such, the time resolution of the set-up for all probe photon energies was $\sim$50 fs.
	
\subsection*{\textit{Ab initio} calculations}
	
To characterize the lattice modes of LCO, we performed density-functional theory linear-response calculations as implemented in the Quantum Espresso package \cite{giannozzi2017advanced}. We used norm-conserving pseudopotentials explicitly including semi-core states \cite{reis2003first} for La and Cu, the local density approximation (LDA) \cite{perdew}, and a plane-wave cutoff energy of 200 Ry on the kinetic energy. The charge density and dynamical matrices were calculated for the $\Gamma$ point of the Brillouin zone using a 7 $\times$ 7 $\times$ 7 $\Gamma$-centered Monkhorst-Pack \cite{monkhorst1976special} electron-momentum grid and a Gaussian smearing of 0.002 Ry. The convergence with respect to all these parameters has been checked thoroughly. The experimental primitive unit cell \cite{takahashi1994structural} was relaxed prior to the phonon calculation which resulted in a slightly reduced volume ($\sim$1\%), typical for LDA calculations.\\ We modelled the optical data of LCO by combining the QS$GW$ theory and DMFT calculations~\cite{swagata_prx}, as implemented in the Questaal package \cite{questaal,questaal-paper}. The paramagnetic DMFT was combined with the QS$GW$ via local Cu-3$d$ projectors of the Kohn-Sham space to the correlated subspace. We carried out the calculations for non-magnetic LCO in the orthorhombic phase with space-group 64/Cmca \cite{structureLA214}, within the QS$GW$+DMFT approach \cite{swag19}. DMFT provides a non-perturbative treatment of the local spin fluctuations. We used hybridization expansion flavor of locally exact continuous time quantum Monte-Carlo solver \cite{KH_ctqmc} to solve the correlated impurity problem. Charge self-consistency on the static QS\emph{GW} potential was performed on a $16 \times 16 \times 16$ \emph{k}-mesh, fully converging the charge.  The \emph{GW} self-energy ($\Sigma^0$), which varies with \emph{k} much more slowly than the kinetic energy, was calculated on a $4 \times 4 \times 4$ \emph{k}-mesh, and converged with RMS change in $\Sigma^0{<}10^{-5}$\,Ry. Subsequent DMFT calculations were iterated, and the dynamical self-energy converged in $\sim$15-20 iterations.  The calculations for the single particle responses were performed with $10^8$ QMC steps per core and the statistics was averaged over 64 cores. As a second step, we considered a small rigid displacement of the ionic positions in LCO along the phonon vector fields calculated from density-functional theory. Finally, we computed the optical conductivity for the different displaced structures. The largest amplitude of the considered shifts were 0.04 \AA.

\section{Acknowledgements}
	
We are grateful to Anthony J.~Leggett, Antoine Georges, Jos\'e Lorenzana, and Ferdi Aryasetiawan for insightful discussions. E.B. and F.C. acknowledge support from the NCCR MUST. T.B. and A.R. were supported by the European Research Council (ERC-2015-AdG694097), European Union H2020 program under GA no.676580 (NOMAD), and Grupos Consolidados (IT578-13). M.A.S. acknowledges support by the DFG through the Emmy Noether programme (SE 2558/2-1). This work was supported by EPSRC (EP/R02992X/1, EP/N02396X/1, EP/M011631/1), and the Simons Many-Electron Collaboration. E.S. and C.B. acknowledge funding from the SNSF by Grant No. 200020-172611. For computational resources, S.A., M.v.S., and C.W. were supported by the ARCHER UK National Supercomputing Service and the UK Materials and Molecular Modelling Hub for computational resources (EPSRC Grant No. EP/ P020194/1); T.B. was supported by the MPCDF Garching.

\section{Author contributions}
E.B. performed the ultrafast optical experiments. E.B., E.S., and F.L. performed the spectroscopic ellipsometry measurements. E.B. and F.C. analyzed the experimental data. S.A., C.W., and M.v.S. performed the QS$GW$+DMFT calculations. T.B., M.A.S., and A.R. performed the phonon calculations. E.P. synthesized the crystals. E.B., M.A.S., C.B., A.R., and C.W. wrote the article. All authors participated in the final version of the article. E.B., M.A.S., C.W., and A.R. conceived the study.

\section{Author information}
The authors declare no competing financial interests. Correspondence and requests for materials should be addressed to E.B. (email: ebaldini@mit.edu), A.R. (email: angel.rubio@mpsd.mpg.de), and C.W. (email: cedric.weber@kcl.ac.uk).
	
\clearpage
\newpage


\setcounter{section}{0}
\setcounter{figure}{0}
\renewcommand{\thesection}{S\arabic{section}}  
\renewcommand{\thetable}{S\arabic{table}}  
\renewcommand{\thefigure}{S\arabic{figure}} 
\renewcommand\Im{\operatorname{\mathfrak{Im}}}
\titleformat{\section}[block]{\bfseries}{\thesection.}{1em}{} 

\section{S1. Steady-state Optical Data}

We simultaneously fitted the measured real ($\epsilon_{1}$) and imaginary ($\epsilon_{2}$) parts of the dielectric function along each polarization channel at 10 K with a Lorentz model. In this approach, Lorentz oscillators are used to describe the contribution of the interband transitions to the dielectric response. The analytical expression of $\epsilon_{1}$ and $\epsilon_{2}$ of the Lorentz functions can be found elsewhere~\cite{jellison2000characterization}. The combined fit of $\epsilon_{1}$ and $\epsilon_{2}$ provides an accurate description of the optical quantities in the energy range of interest and it is crucial to overcome the issues related to a Kramers-Kronig analysis of the transient reflectivity data. The results of the fit are shown in Fig. S1 $A,B$ as solid lines. Along the $a$-axis, we find that the relevant oscillators contributing to the spectral region over which our pump-probe experiment is sensitive are centered at $a_{1}$ = 1.73 eV, $a_{2}$ = 2.16 eV and $a_{3}$ = 2.64 eV. Oscillator $a_{1}$ well agrees with the energy of a $d$-$d$ crystal field excitation reported by optical spectroscopy \cite{ref:falck, mcbride_ellipsometric_1994} and resonant inelastic x-ray scattering \cite{ref:ellis}. Oscillator $a_{2}$ is the excitation across the optical CT gap, whereas oscillator $a_{3}$ represents the high-energy shoulder to the optical CT excitation \cite{ref:falck, mcbride_ellipsometric_1994, ref:ellis}. Over the same energy region along the $c$-axis, Lorentz oscillators are instead centered at $c_{1}$ = 1.49 eV and $c_{2}$ = 4.08 eV.

\section{S2. Temperature Dependence of the Optical Properties}

In this Section, we show how the optical properties of LCO respond to an increase in the lattice temperature. This analysis allows to identify any thermal response in the pump-probe experiment.

First, we estimate the maximum transient lattice temperature that is reached in a crystal of LCO upon photodoping with the maximum excitation density used in our pump-probe experiment. To this aim, we rely on the simple expression
\begin{equation}
Q = \int_{T_i}^{T_f} mC(T) \delta T
\end{equation}
where $Q$ is the absorbed heat from a single laser pulse, $m$ is the illuminated mass, $C(T)$ is the temperature-dependent specific heat, $T_i$ = 10 K is the initial equilibrium temperature, and $T_f$ is the final temperature. We use our photoexcited carrier density values and calculate $m$ through the material density $\rho$ = 6.92 g/cm$^3$ and the illuminated sample volume $V$ = 4 $\times$ 10$^{-9}$ cm$^3$. We consider the temperature dependence of the heat capacity, as measured at low temperature in Ref. \cite{jorge2004thermodynamic} and extrapolated to high temperature with a linear behavior. At the minimum excitation density used in the experiments the calculation yields $T_{f,min}$ = 79 K, whereas at the maximum excitation density we obtain $T_{f,max}$ = 128.5 K. The latter temperature is still well below the N\'eel temperature of our sample ($T_N$ = 307 K).\\

As a second step, we measure the temperature dependence of the optical response via spectroscopic ellipsometry along the $a$- and $c$-axis between 10 K and 300 K. The data are shown in Figs. S2 $A-F$. In the following, we restrict the visualization of the data only to the spectral region that is relevant to our pump-probe experiment. An interpolation procedure has been applied to increase the number of data points and improve the presentation of the changes induced in the optical spectra by the temperature increase.

The real and imaginary part of the $a$-axis optical conductivity are shown in Figs. S2 $A,B$ at 10 K and 150 K. We observe that the optical CT excitation in $\sigma_{1a}$ broadens and redshifts with increasing temperature, in agreement with previous studies \cite{ref:falck}. The inset of Fig. S2 $A$ zooms in the evolution of the optical CT feature close to the peak energy: The crossing point between each of the high-temperature traces and that at 10 K also redshifts with increasing temperature, as indicated by the arrows. As a result, the temperature-driven differential conductivity $\Delta\sigma_{1a}$ (brown curve in Fig. S2 $C$) shows a positive feature at lower energy, followed by a negative one at higher energy. This is again consistent with previous reports \cite{ref:okamoto_2, novelli2014witnessing}. The same temperature increase also influences $\sigma_{2a}$ (which is $<$ 0), yielding a sizable decrease of its absolute value over the 1.90-2.30 eV energy range. In this case, the differential conductivity $\Delta\sigma_{2a}$ (violet curve in Fig. S2 $C$) results in a positive contribution.

We now focus on the $c$-axis optical response. Here, since the optical quantities are rather temperature independent in the 10-150 K range, we compare the response at 10 K with that at 300 K. This allows us to enhance the contrast to visualize the temperature-induced variation of the optical quantities. In Fig. S2 $D$, we observe that $\sigma_{1c}$ is slightly reduced with increasing temperature, which implies a negative $\Delta\sigma_{1c}$ (brown curve in Fig. S2 $F$). Furthermore, the absolute value of $\sigma_{2c}$ (Fig. S2 $D$) increases over the the 1.90-2.50 eV range. The resulting $\Delta\sigma_{2c}$ has a negative sign (violet curve in Fig. S2 $F$).

\section{S3. Estimation of the Time Resolution in the Pump-probe Data}

We also estimate the temporal resolution of our ultrafast experiments by identifying a resolution-limited rise in the $\Delta$R/R temporal traces. This is shown in Fig. S3, where the temporal trace has been selected around 2.60 eV in the $c$-axis probe polarization channel. We observe an ultrafast rise of $\sim$50 fs (peak-to-peak) in the transient response, which is a clear fingerprint of the time-resolution of our experimental setup. The same apparatus was used in several studies to generate and detect coherent optical phonons (with energies as high as 70 meV) in other materials \cite{mann2015probing, mann2016probing, baldini2017clocking, borroni2017coherent, baldini2018lattice}.

\section{S4. Comparison with Other Materials}

In this Section, we show that the ultrafast response observed in the rise of the LCO temporal traces is an intrinsic property of this material and not an artifact produced by our experimental set-up. To this aim, we measure other samples under the same experimental conditions, namely a thin film of the anisotropic metal MgB$_2$ and a single crystal of the band semiconductor CH$_3$NH$_3$PbBr$_3$. We excited both materials with a 45 fs laser pulse centered around 3.10 eV and monitored the pump-probe response with the same time step utilized to measure LCO ($\sim$ 13 fs). The $\Delta$R/R temporal response from the different materials is shown in Fig. S4. Here for simplicity we have selected the temporal traces around a photon energy of 2.45 eV, but the results are general and apply to the whole measured spectral range and for different pump fluences. We observe that only the temporal traces of LCO show a resolution-limited signal before 0.17 ps, whereas the rise time in MgB$_2$ and CH$_3$NH$_3$PbBr$_3$ is significantly slower and corresponds to the results reported in the literature with similar excitation schemes \cite{baldini_MgB2}. These results rule out the involvement of a material-independent optical nonlinearity in the ultrafast signal of LCO.

\section{S5. Transient Reflectivity}

Figures S5 $A,C$ display the color-coded maps of the $\Delta$R/R response as a function of the probe photon energy and of the time delay between pump and probe at 10 K, for a probe polarization along the $a$- and $c$-axis, respectively. In both cases, the pump polarization is set along the $a$-axis and the excitation density is estimated at 0.06 photons/Cu. The $a$-axis $\Delta$R/R response (Fig. S5 $A$) comprises a negative feature centered around 2.15 eV, which corresponds to the optical CT excitation. This response is consistent with that of previous studies of the in-plane charge dynamics \cite{novelli2014witnessing, mann2015probing}. Unlike the signal obtained upon below-gap excitation, here the response is one order of magnitude larger due to the enhanced absorption of particle-hole pairs across the optical CT gap. In addition, the high-energy region of the spectrum displays a sign inversion around 0.5 ps, as evident from the time traces in Fig. S5 $B$.

The $c$-axis $\Delta$R/R response in Fig. S5 $C$ shows instead a more complex behavior, with two distinct regions of negative and positive photoinduced changes. These features are well displayed by selecting specific temporal traces across the probed spectral range, as evidenced in Fig. S5 $D$. Again, an extremely fast signal appears in the rise of the response during the first hundred of femtoseconds. The subsequent relaxation dynamics comprises a multi-exponential decay. On top of this background, a prominent oscillatory pattern clearly emerges across the whole spectrum.

To unravel the degree of anisotropy governing the $a$- and $c$-axis response, we compare two temporal traces selected from the color-coded maps of Figs. S5 $A,C$. The results are shown in Fig. S6 $A,B$, in which the traces have been normalized for comparison. It is evident that the $c$-axis relaxation dynamics is faster than its $a$-axis counterpart, which represents a signature of a more insulating behavior shown by the out-of-plane charge transport. We also observe that, in the in-plane dynamics, the high signal intensity resulting from the optical CT excitation hides a clear manifestation of the coherent oscillations. In contrast, in the out-of-plane dynamics, a beating among several coherent modes is clearly distinguished and found to persist up to several picoseconds. Despite this difference in the relaxation, the rise time of both traces is found to be identical within our experimental resolution, as shown in Fig. S6 $B$. This indicates that the same phenomenon governs the rise time along the two crystallographic axes of LCO.\\

\section{S6. Transient Optical Conductivity}

From our nonequilibrium experiment we extract the transient complex optical conductivity $\Delta\sigma(\omega, t)$ = $\Delta\sigma_1$($\omega$, t) + i $\Delta\sigma_2$($\omega$, t). This can be calculated without the need of a Kramers-Kronig analysis by relying on our steady-state spectroscopic ellipsometry data of Fig. S1 as a starting point and performing a differential Lorentz model of the $\Delta$R/R maps. This method has been previously utilized to treat ultrafast broadband optical data \cite{mansart2012evidence, novelli2012ultrafast} and it represents the most accurate procedure currently developed.

\begin{figure*}[ht!] 
	\centering
	\includegraphics[width=\columnwidth]{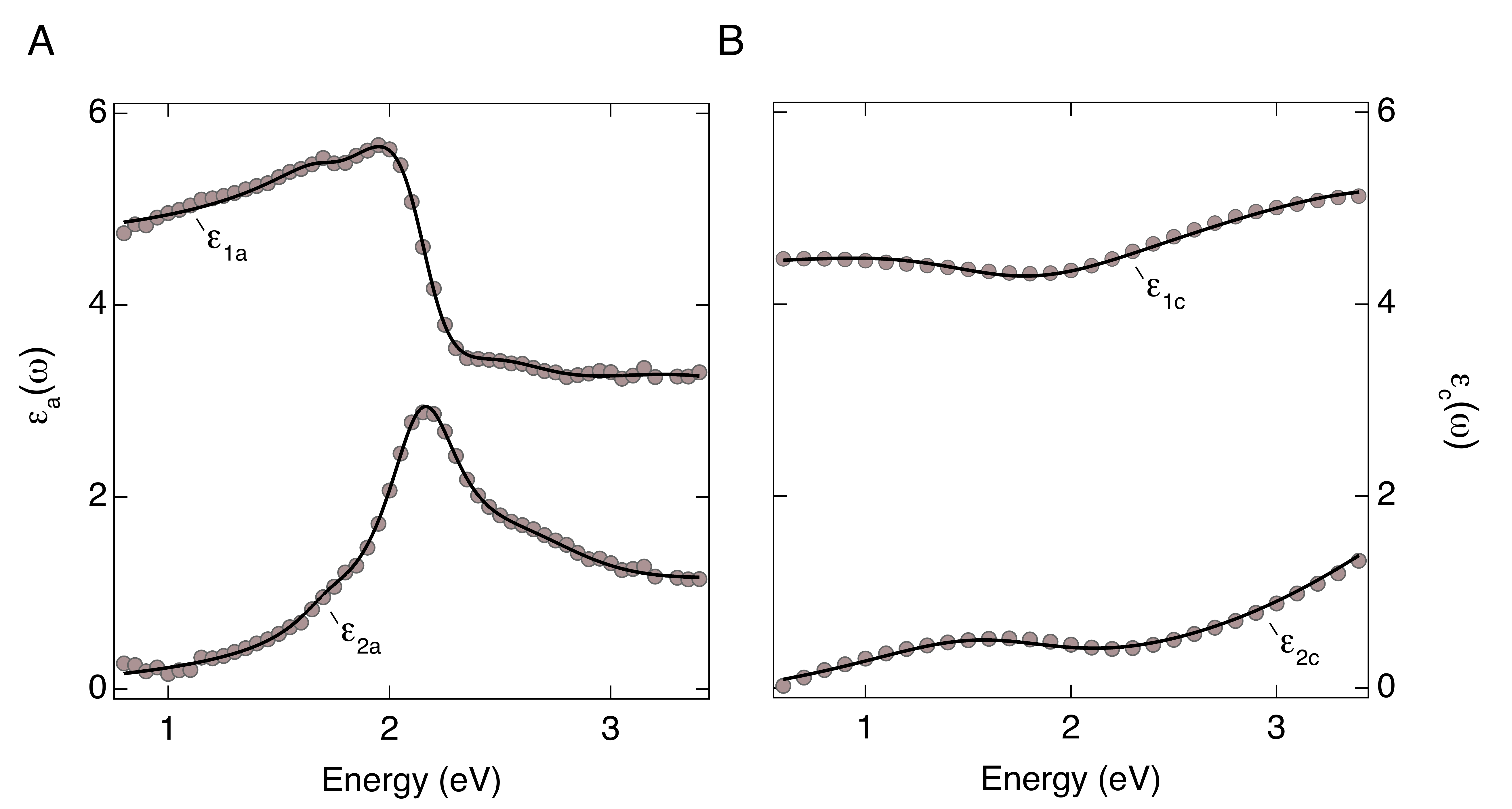}
	\caption{\textit{(A,B)}, Real ($\epsilon_1$($\omega$)) and imaginary ($\epsilon_2$($\omega$)) parts of the complex dielectric function of LCO, measured at 10 K with spectroscopic ellipsometry (brown dots) along the ($A$) $a$-axis and ($B$) $c$-axis. Solid lines show fits of the measured data to a Lorentz model.}
	\label{Epsilon}
\end{figure*}

To extract $\Delta\sigma(\omega,t)$ from the $\Delta$R/R data, we proceed as follows. We use the static reflectivity $R$ determined via spectroscopic ellipsometry and obtain the momentary reflectivity $R(\omega,t)$ in the range explored by the pump-probe experiment by multiplying $\Delta R/R(\omega,t)$ at a fixed time delay $t$ by $R$ itself. As starting parameters, we use those describing the steady-state response of Fig. \ref{Epsilon}. Letting one (two) Lorentz oscillator(s) free to vary in the frequency range along the $a$-axis ($c$-axis) is sufficient to accurately reproduce the $R(\omega,t)$ spectra at all times delays. In a final step, the transient conductivity spectra $\Delta\sigma(\omega, t)$ = $\Delta\sigma_1$($\omega$, t) + i$\Delta\sigma_2$($\omega$, t) are recombined into the full maps of Fig. \ref{DeltaSigma}.

\noindent The long temporal window covered by our experiment at high time resolution allow us to refine the model of Okamoto \textit{et al.}~\cite{ref:okamoto_2} and extract information about the temporal response along the $a$- and $c$-axis. We find that both responses can be fitted only with Eq. (2):
\begin{equation}
\begin{split}
f(t) = C_G~\textup{exp} \Bigg( \frac{-(t-D)^2}{\tau_{R_1}} \Bigg) + \sum_{i} C_i \int_{-\infty}^{t} \textup{exp} \Bigg(- \frac{t-t^{\prime}}{\tau_{i}} - \frac{(t-D)^2}{\tau_{R_1}}\Bigg)dt^{\prime} +\\
+ C_H \int_{-\infty}^{t} \Bigg[1-\textup{exp}\Bigg(-\frac{t-t^{\prime}}{\tau_{R_2}}\Bigg)\Bigg]\textup{exp}\Bigg(-\frac{t-t^{\prime}}{\tau_{H}} - \frac{(t-D)^2}{\tau_{R_2}}\Bigg)dt^{\prime}
\end{split}
\end{equation}

\noindent The first term describes a pulsed response that captures the fast dynamics related to the metallic state. The second term comprises distinct exponential relaxation processes ($i$ = 3 for the in-plane response and $i$ = 2 for the out-of-plane response) due to the decay of the metallic state and the charge localization in mid-gap states. The last delayed component corresponds to the bolometric (heating) response of the sample, which sets after the thermalization of the excited carriers has occurred. $\tau_{R_1}$ = 0.01 $\pm$ 0.001 ps is the rise time of the Gaussian term, whereas $\tau_{R_2}$ = 0.03 $\pm$ 0.0003 ps is the rise time of the exponential components.; $\tau_{i}$ are the relaxation constants of the exponential decays; $D$ is a delay parameter with respect to the zero time. The results of the fit are shown in Fig. \ref{GlobalFit} $A,B$ as dashed black lines superposed on the original data. The timescales retrieved along the \textit{a}-axis are $\tau_{1}$ = 0.06 $\pm$ 0.002 ps, $\tau_{2}$ = 0.4 $\pm$ 0.003 ps, $\tau_{3}$ = 1 $\pm$ 0.008 ps, and $\tau_{H}$ = 0.65 $\pm$ 0.07 ps, while those along the \textit{c}-axis are $\tau_{1}$ = 0.06 $\pm$ 0.004 ps, $\tau_{2}$ = 1.9 $\pm$ 0.01 ps, and $\tau_{H}$ = 3 $\pm$ 1.85 ps. The fact that both fits along the $a$- and $c$-axis indepentently provide the same values for $\tau_{R_1}$ and $\tau_{R_2}$ strongly supports the idea (already evident from the inspection of the raw data) that the same ultrafast phenomenon (namely the ultrafast metallic state) emerges along both crystallographic directions.

\section{S7. Photon Energy Dependence of the Collective Modes' Amplitudes}

In this Section, we benefit from the use of a broadband continuum probe to extract the photon energy dependence of the oscillation amplitude for four distinct Raman-active modes in the pump-probe experiment. To this aim, we focus on the $\Delta$R/R response along the material's $c$-axis, along which the oscillations emerge clearly. We select twenty temporal traces from the map displayed in Fig. \ref{DeltaReflectivity} $C$ and perform a global fit analysis by imposing the same relaxation time constants across the monitored spectral region. By calculating the Fourier transform of the residuals, for each collective mode we reconstruct its intensity profile, as shown in Fig. \ref{Raman}. The low intensity of mode $A_g(5)$ prevented us from extracting its photon energy dependence. Similar results were obtained by decreasing the excitation density to 0.023 ph/Cu, which is still above the threshold to induce the metallic state in LCO \cite{ref:okamoto_2}.

As evident from Fig. \ref{Raman}, the oscillatory response of the in-plane pump/out-of-plane probe data is dominated by the optical phonon $A_g$(3), which corresponds to the in-phase vibration between La and apical O atoms. The intensity of this mode shows a first feature around 2.00 eV, followed by a marked increase above 2.50 eV. The strong enhancement of the mode intensity towards high energies agrees with the idea that the $c$-axis electrodynamics in the 2.60-3.00 eV spectral region is dominated by transitions from Cu to apical O states \cite{ref:uchida_LSCO, lorenzana2013investigating, ref:uchida_caxis}. Interestingly, this mode was also found to oscillate in the $c$-axis $\Delta$R/R response of optimally-doped La$_{1.85}$Sr$_{0.15}$CuO$_4$ upon in-plane photoexcitation at 1.55 eV \cite{mansart2013coupling, lorenzana2013investigating}. Consistent with our measurements, also in La$_{1.85}$Sr$_{0.15}$CuO$_4$ this phonon was found to resonate at an energy scale above 2.50 eV.

Mode $A_g(4)$, corresponding to the vibrations of the in-plane O along the $c$-axis, produces a sizable effect in the region between 1.80 and 2.50 eV and resonates around 2.00 eV. This resonant enhancement around 2.00 eV allows us to conclude that the underlying interband transition is modulated by displacements along the $c$-axis of the material. Thus, the origin of this interband transition may lie in the transfer of a hole from a Cu atoms to a planar O of a contiguous plane. The estimate of the Madelung energy for such a process ($\sim$ 2.00 eV) supports this scenario \cite{lorenzana2013investigating}.

The complete mapping of the Raman profiles along the $c$-axis opens perspectives towards the evaluation of the electron-phonon matrix elements for all $A_g$ modes of LCO in future \textit{ab initio} calculations. 

\section{S8. Additional QS$GW$+DMFT Results}

The compound with undisplaced unit cell has been recently characterized by the local density approximation (LDA) and the QS$GW$+DMFT methods (see Ref.~\cite{swagata_prx}). The LDA bands at the Fermi level consist of strongly mixed (3:2) Cu-3$d_{x^{2}-y^{2}}$ and O-2$p_{x}, p_{y}$ orbitals. The Cu-3$d_{z^{2}}$ orbitals also have strong mixing with primarily apical O-2$p_{z}$ orbital at 1.90 eV below the Fermi level. The O-2$p$ states are strongly hybridized with most orbital characters throughout the entire window $[E_{F}{-}8\mathrm{\,eV},E_F]$. However, unlike in LDA, the self-energy in QS\emph{GW} (the off-diagonal, dynamic and momentum dependent matrix elements) splits out the O-2$p$ orbitals from the Cu-3$d$ ones and puts them roughly 2 eV below the band with dominant Cu 3$d_{z^{2}}$ character.  Within the QS\emph{GW} description, the band character at the Fermi level becomes predominantly Cu-3$d_{x^{2}{-}y^{2}}$, with significantly lessened O-2$p_{x}, p_{y}$ contribution than in LDA. In particular, the QS\emph{GW} bands closest to the Fermi level consist of 2:1 admixtures of Cu-3$d_{x^{2}{-}y^{2}}$ and O-2$p_{x},p_{y}$ orbitals. However, we find that the QS\emph{GW} band at the Fermi level has finite but small contribution from the Cu-4$s$ orbital on the $A{-}C_{0}{-}E_{0}$ and $S{-}R{-}Z$ symmetry lines. Its weight is more pronounced near the Y point, in a band with Cu-4$s$ character. O-2$p_{z}$ below -3 eV hybridize to some extent with Cu-3$d_{z^{2}}$ around -2 eV and the other Cu-3$d$ orbitals, and to a lesser extent, the bands at $E_{F}$.

The main advantage of non-magnetic QS$GW$ bands for the displaced structure of LCO is to get rid of the $f$-states (otherwise present in density-functional theory calculations) from the spectral region close to Fermi energy. Furthermore, the off-diagonal orbital elements of QS$GW$ self-energy makes the Cu-3$d_{x^2-y^2}$ character prominent in the active band at Fermi energy and it also controls the relative orientations of different O and Cu-3$d_{z^2}$ states with respect to the Cu-3$d_{x^2-y^2}$ ones \cite{swagata_prx}. Relative orientations of different Cu-3$d$ states and in-plane O-$2p$ and apical O-$2p$ states are tabulated in Table I. On analyzing the bare energies for different orbitals presented in the Table I, we find that there is no clear trend in the change in CT energies that distinguishes the B$_{g}$ modes from the A$_{g}$ modes. On average, all Raman-active modes simulate a change in in-plane and out-of-plane CT energies by $\sim$30-40 meV. However, the electronic structure for LCO displaced along the A$_{g}$(3) mode coordinates is characterized by an apical O-$p_{z}$ state that is nearly degenerate with the Cu-3$d_{z^2}$. In the electronic structure for LCO displaced along the A$_{g}$(5) mode eigenvector, the Cu-3$d_{x^2-y^2}$-3$d_{z^2}$ splitting reduces by $\sim$130 meV and the in-plane CT energy drops by $\sim$70 meV. Moreover, the apical O-$p$ states are fairly closer to the $d_{x^2-y^2}$ state in the electronic structure of LCO displaced along the A$_{g}$(2) and A$_{g}$(5) modes. Finally, in the electronic structure obtained upon displacements along the B$_{1g}$ and B$_{2g}$ modes, the in-plane CT energy increases by $\sim$30-40 meV. 

Once QS$GW$ is coupled with DMFT, the local dynamical spin fluctuations put the system on the brink of a Mott transition. Irrespective of whether the modes are A$_{g}$ or B$_{g}$, the low-energy properties can not be characterized in a purely Bloch band-like scenario. When the unit cell is displaced along the $B_g$ modes eigenvectors, the system appears to evolve through a local dynamic IMT with an imaginary part of the local self-energy that picks up a pole at low energies $\omega\sim 0$, effectively suppressing the low-energy quasiparticle excitation. However, in case of B$_{3g}$ mode, incoherent excitations can be observed across Fermi energy, while quasi-particle like excitations are absent. An incoherent metal with large effective mass characterizes the electronic structure of the system displaced along the $A_g$ modes coordinates. Displacements along A$_{g}$(1) lead to a  single-particle response that is similar to the one for the compound displaced along the B$_{3g}$ mode. However, for other A$_{g}$ modes at low energy it is still possible to perform low-energy quasiparticle fittings. In these cases, the QS$GW$+DMFT electronic spectral functions show a three-peak structure that is typically found in the strong coupling regime of correlated electron systems; both the incoherent, heavy quasiparticle excitation at low energy and atomic-like Hubbard ``band" features at high energy can be observed. The lower Hubbard band is extremely broad and spreads over $\sim$6-8 eV, in excellent agreement with the data reported in Ref. \cite{nucker}.  The bad metallic nature of the broad incoherent quasiparticle excitation present at $\omega$=0 can be characterized for the different $A_g$ modes by computing the mass enhancement factor ($m^*/m$) and single-particle scattering rate ($\gamma$). The results are listed in Table II. In the last column of Table II, we show the effective doping concentration in the Cu-3$d$ states when the unit cell of LCO is displaced along the different Raman-active modes. We observe that the modes that provide more metallicity simulate a stronger effective doping (hole doping in case of A$_g$ modes and electron doping in case of B$_{3g}$ mode) in the Cu-3$d$ states. However, the total electron count of all Raman-active modes are same, and the systems are not doped in reality. The effective doping in B$_{1g}$ and B$_{2g}$ modes is almost negligible in comparison to undisplaced LCO.  
\newpage
\clearpage

\begin{table}
	\begin{center}
		\footnotesize
		\begin{tabular}{ccccccccc}
			\hline
			Mode & Cu-3$d_{x^2-y^2}$ & Cu-3$d_{z^2}$ & apical O-2$p_{y}$ & AO-2$p_{z}$ &  AO-2$p_{x}$ & IO-2$p_{y}$ & IO-2$p_{z}$ &  IO-2$p_{x}$\\
			\hline
			Undisplaced  &  0.000000 & -0.729280 & -0.212671 & -0.493781 & -0.199210 & -1.486084 & -0.929154 & -1.484877 \\      
			
			A$_g$(1)  &  0.000000 & -0.707354 & -0.261992 & -0.481159 & -0.228354 & -1.485348 & -0.921401 &  -1.477706 \\
			
			A$_{g}$(2)  &  0.000000 & -0.751029 & -0.154252 & -0.452560 & -0.156449 & -1.519043 & -0.932744 & -1.508519 \\
			
			A$_{g}$(3)  &  0.000000 & -0.718872 & -0.400241 & -0.719047 & -0.399869 & -1.483198 & -0.928412 & -1.479216 \\
			
			A$_{g}$(4)  &  0.000000 & -0.684008 & -0.239187 & -0.534775 & -0.206341 & -1.488909 & -0.916818 & -1.486509 \\
			A$_{g}$(5)  &  0.000000 &  -0.591483 & -0.086867 &  -0.364024 &  -0.079538 &  -1.415804 &  -0.851139 &  -1.413909 \\
			
			B$_{1g}$  &  0.000000 & -0.681711 & -0.253692 & -0.513599 &  -0.248558 & -1.509005 & -0.948838 &  -1.513572           \\
			
			B$_{2g}$  &  0.000000 & -0.657151 & -0.315573 & -0.562927 & -0.277567 & -1.534287 &  -0.978037 &  -1.530722           \\
			
			B$_{3g}$  &  0.000000 & -0.686722 & -0.184135 & -0.460328 & -0.165544 & -1.494214 & -0.920689 & -1.494236            \\

		\end{tabular}
	\end{center}
	\caption{Relative energies of different orbitals with respect to the Cu-3$d_{x^2-y^2}$. All numbers are given in eV. AO = apical oxygen; IO = in-plane oxygen.}
\end{table}
\clearpage
\newpage

\begin{table}
	\begin{center}
		\footnotesize
		\begin{tabular}{ccccccc}
			\hline
			Mode & $m^*/m$ & $\gamma$ (meV) & Effective Cu-3$d$ doping ($\delta$) \\
			\hline
			$A_g$(1) & 24 & 100  & 0.7\% \\
			
			$A_g$(2) & 19 & 26  & 3.5\% \\
			
			$A_g$(3) & 24 & 15  & 2.7\%  \\
			
			$A_g$(4) & 13 & 21  & 3.4\% \\
			
			$A_g$(5)	& 12 & 11  & 3.3\% \\
			
			$B_{1g}$ &  &   & 0.4\% \\
			
			$B_{2g}$ &  &   & 0.3\% \\
			
			$B_{3g}$ &  &   & -1.6\% \\
		\end{tabular}
	\end{center}
	\caption{Quasiparticle mass enhancement factor ($m^*/m$), scattering rate ($\gamma$)  for the metallic state resulting from the displacement of the unit cell of LCO along different $A_g$ and $B_g$ mode coordinates. Effective Cu-3$d$ level hole doping (negative sign for B$_{3g}$ mode implies electron doping) are presented in the last column.}
\end{table}
\newpage
\clearpage

\newpage
\clearpage

\begin{figure*}[ht!] 
	\centering
	\includegraphics[width=\columnwidth]{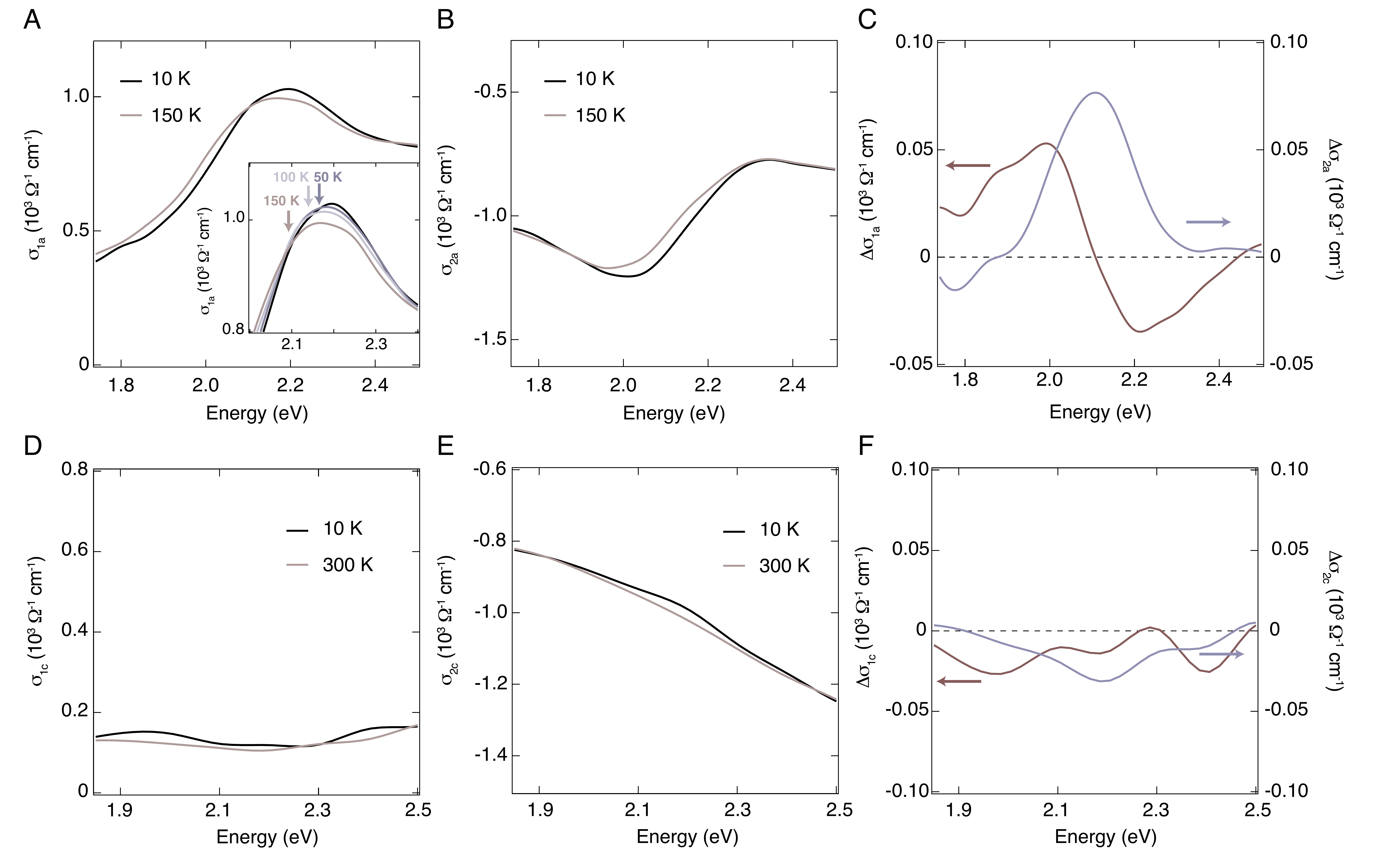}
	\caption{Temperature dependence of the real ($\sigma_{1}$($\omega$)) and imaginary ($\sigma_2$($\omega$)) parts of the complex optical conductivity of LCO. \textit{(A,B)}, Optical response along the $a$-axis (inset of Panel (a) zooms in the region of the optical CT feature) and \textit{(C)}, Differential optical conductivity along the $a$-axis when the temperature is varied from 10 K to 150 K ($\Delta\sigma_a$ = $\sigma_a$(150 K) - $\sigma_a$(10 K)). \textit{(D,E)}, Optical response along the $c$-axis and \textit{(F)} Differential optical conductivity along the $c$-axis when the temperature is varied from 10 K to 300 K ($\Delta\sigma_c$ = $\sigma_c$(300 K) - $\sigma_c$(10 K)).}
	\label{StaticTemperature}
\end{figure*}
\newpage
\clearpage

\begin{figure*}[ht!] 
	\centering
	\includegraphics[width=0.7\columnwidth]{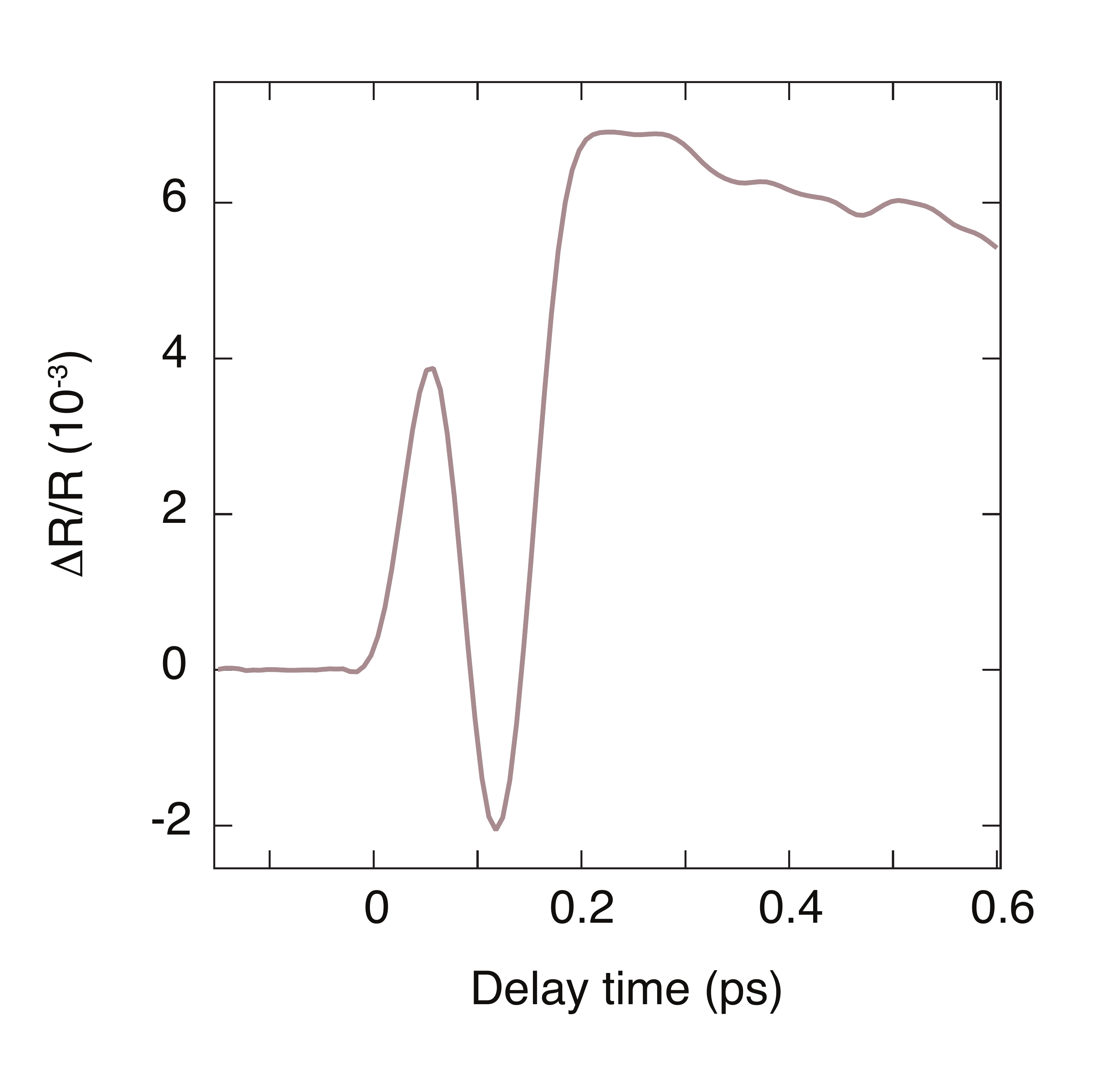}
	\caption{Estimate of the temporal resolution of the pump-probe experiment. The temporal trace shown is selected at a photon energy of 2.60 eV in the experiment with pump polarized along the $a$-axis and probe polarized along the $c$-axis. The rise of the trace is resolution-limited and the time resolution can be estimated around $\sim$0.05 ps.}
	\label{TemporalResolution}
\end{figure*}
\newpage
\clearpage

\begin{figure*}[ht!] 
	\centering
	\includegraphics[width=0.7\columnwidth]{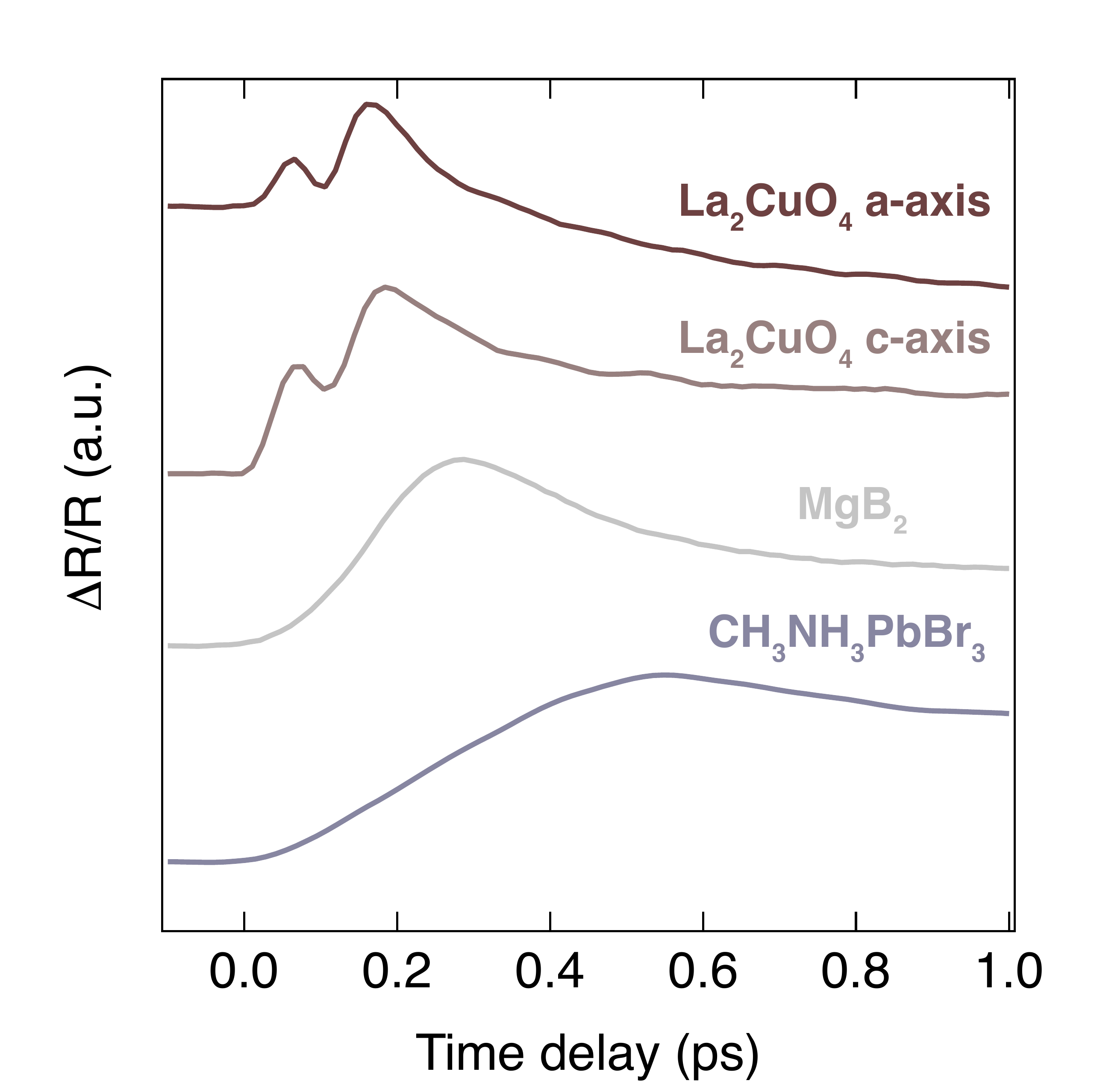}
	\caption{Normalized $\Delta$R/R data measured on different materials photoexcited with 45 fs laser pulses at 3.10 eV. The temporal traces were integrated over 0.10 eV around the probe photon energy of 2.45 eV, but the results are general and apply to the whole measured spectral range ($\sim$1.74-2.60 eV). The comparison is made between LCO single crystals with probe polarization along the $a$- and $c$-axis (brown curves), MgB$_2$ thin films (green curve) and CH$_3$NH$_3$PbBr$_3$ single crystals (violet curve). Only LCO responds to photoexcitation with an extremely fast signal during the first 0.1 ps, indicating that this is an intrinsic material property. For clarity, the LCO response along the $a$-axis has been multiplied by -1.}
	\label{ComparisonMaterials}
\end{figure*}
\newpage
\clearpage

\begin{figure*}[ht!] 
	\centering
	\includegraphics[width=\columnwidth]{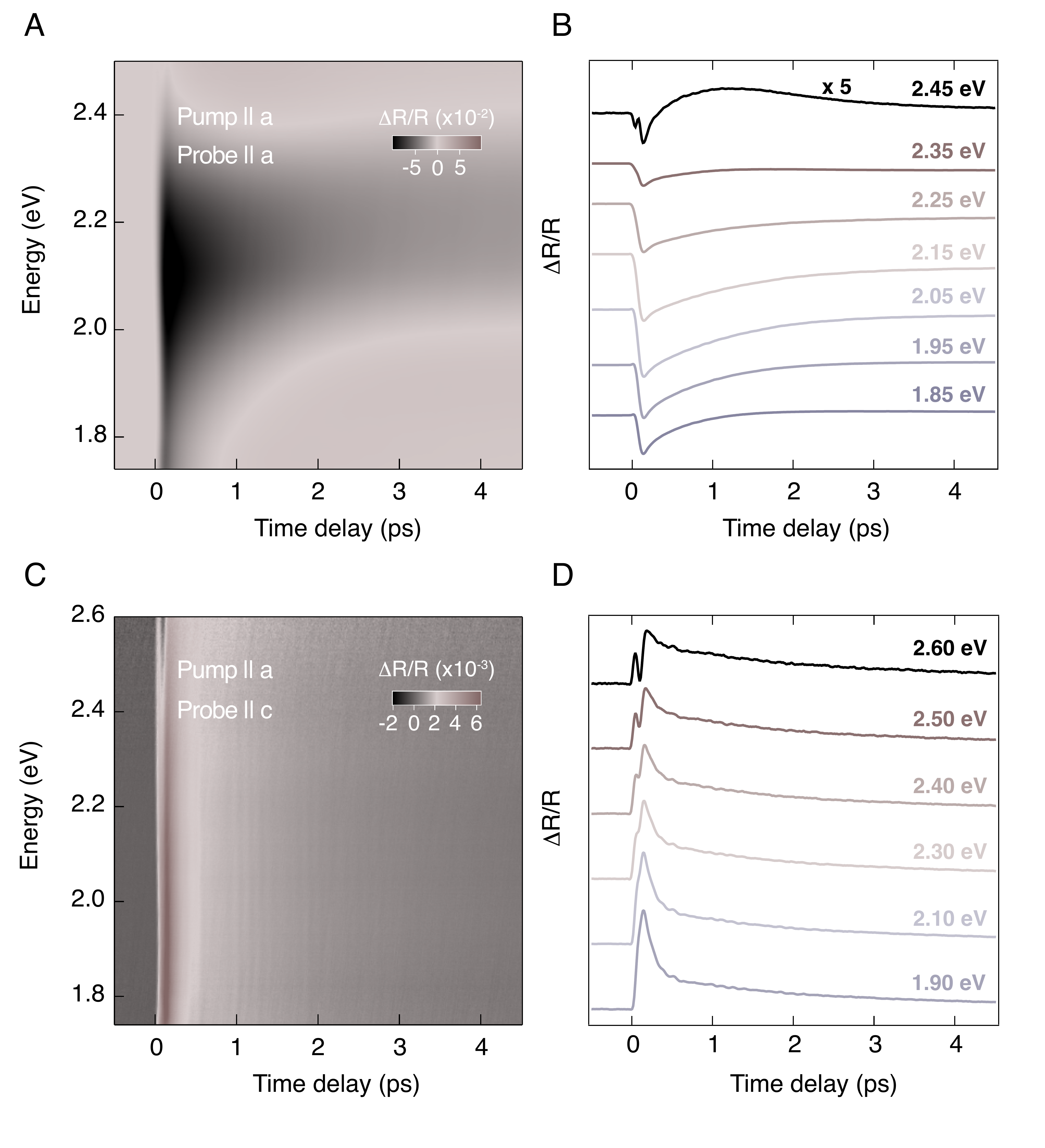}
	\caption{(\textit{A,C}) Color-coded maps of $\Delta$R/R at 10 K with in-plane pump polarization and ($A$) in-plane, ($C$) out-of-plane probe polarization, as a function of probe photon energy and pump-probe time delay. The pump photon energy is 3.10 eV and the excitation density is 0.06 photons/Cu. \textit{(B,D)} Temporal traces at specific probe photon energies of the respective $\Delta$R/R maps. Each temporal trace results from the integration over 0.10 eV around the indicated probe photon energy.}
	\label{DeltaReflectivity}
\end{figure*}
\newpage
\clearpage

\begin{figure*}[ht!] 
	\centering
	\includegraphics[width=\columnwidth]{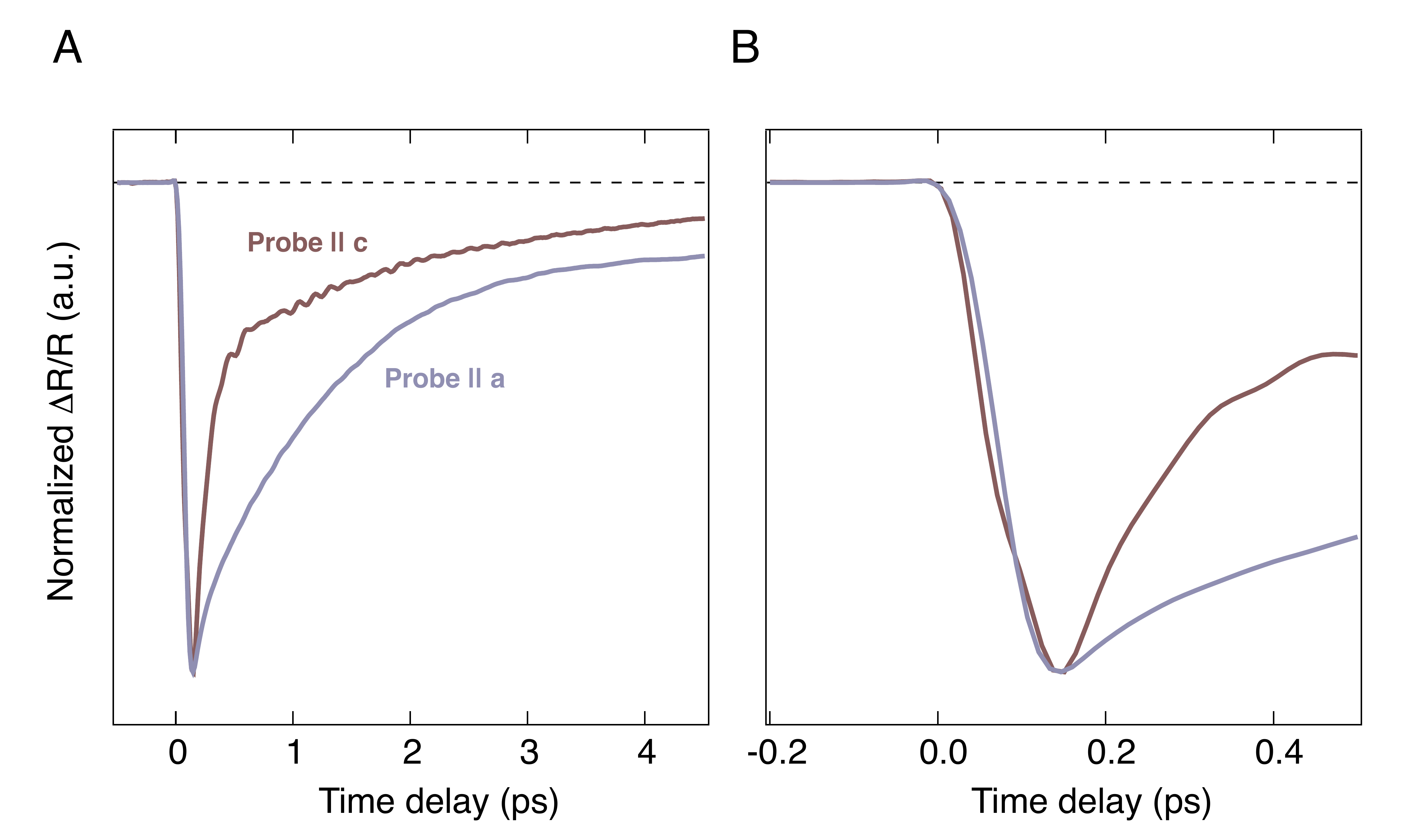}
	\caption{Comparison between the in-plane (violet trace) and out-of-plane (brown trace) temporal evolution of $\Delta$R/R upon in-plane photoexcitation. The traces have been normalized to enable comparison and the sign of the out-of-plane response has been reversed. To maximize the signal-to-noise ratio, the in-plane temporal trace has been selected at a photon energy of 2.10 eV and integrated over 0.20 eV, whereas the out-of-plane trace is cut at 2.00 eV and integrated over 0.40 eV. The excitation density is 0.06 photons/Cu. \textit{(A)}, Complete temporal response. \textit{(B)}, Zoom into the rise time of the response.}
	\label{ComparisonReflectivity}
\end{figure*}
\newpage
\clearpage

\begin{figure*}[ht!] 
	\centering
	\includegraphics[width=0.7\columnwidth]{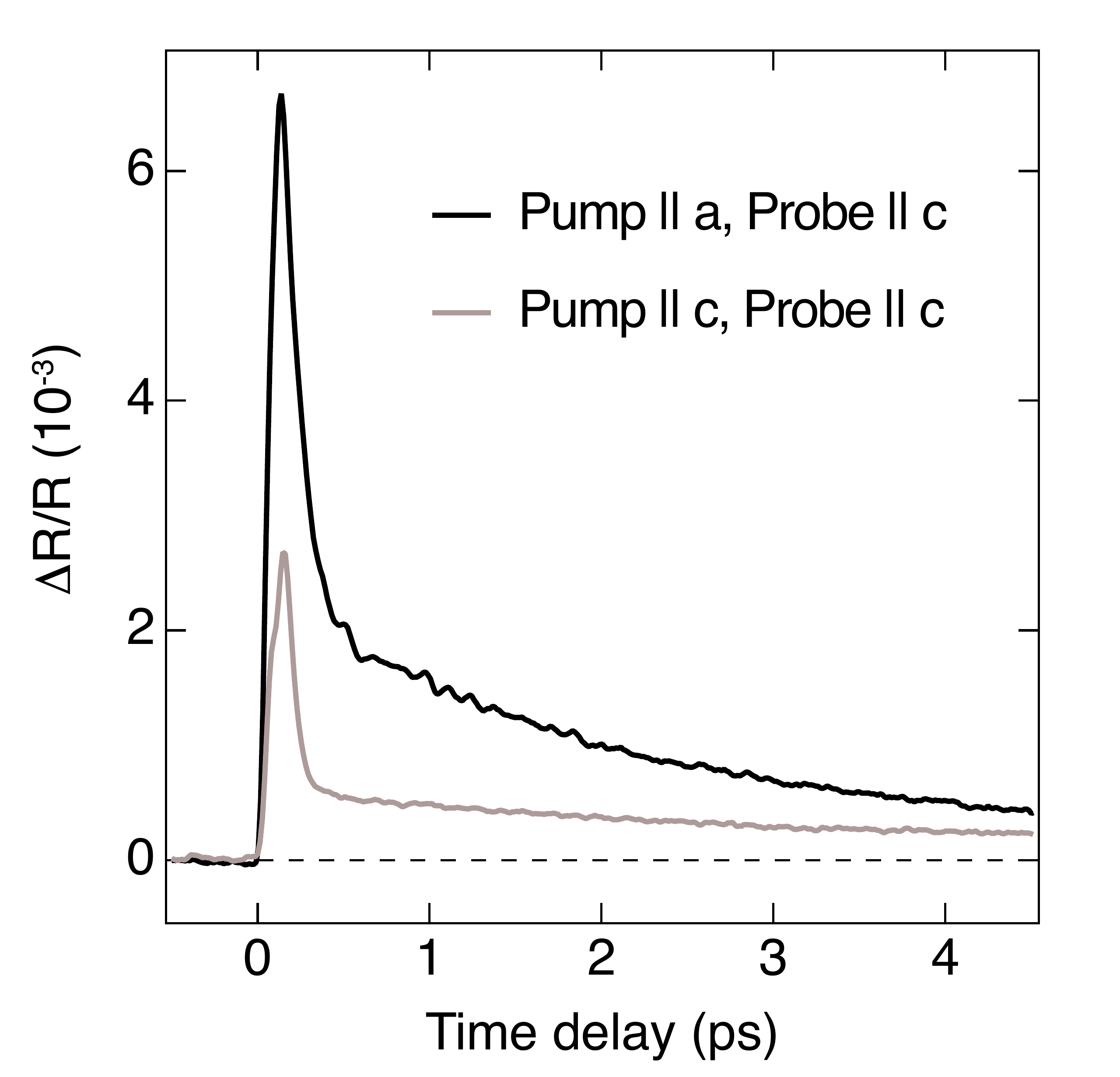}
	\caption{Out-of-plane $\Delta$R/R temporal response for a pump polarization set along the $a$-axis (black curve) and $c$-axis (brown curve). The carrier excitation density is kept constant between the two excitation schemes (0.06 photons/Cu) by relying on the steady-state optical response of LCO. Both traces are obtainted from the integration over 0.10 eV around a photon energy of 1.80 eV. We observe that the signal resulting from out-of-plane photoexcitation has a weaker amplitude than the one obtained from in-plane photoexcitation and the signature of the coherent phonon modes also changes dramatically between the two pump polarization configurations.}
	\label{Pump_c_Probe_c}
\end{figure*}
\newpage
\clearpage



\begin{figure*}[ht!] 
	\centering
	\includegraphics[width=\columnwidth]{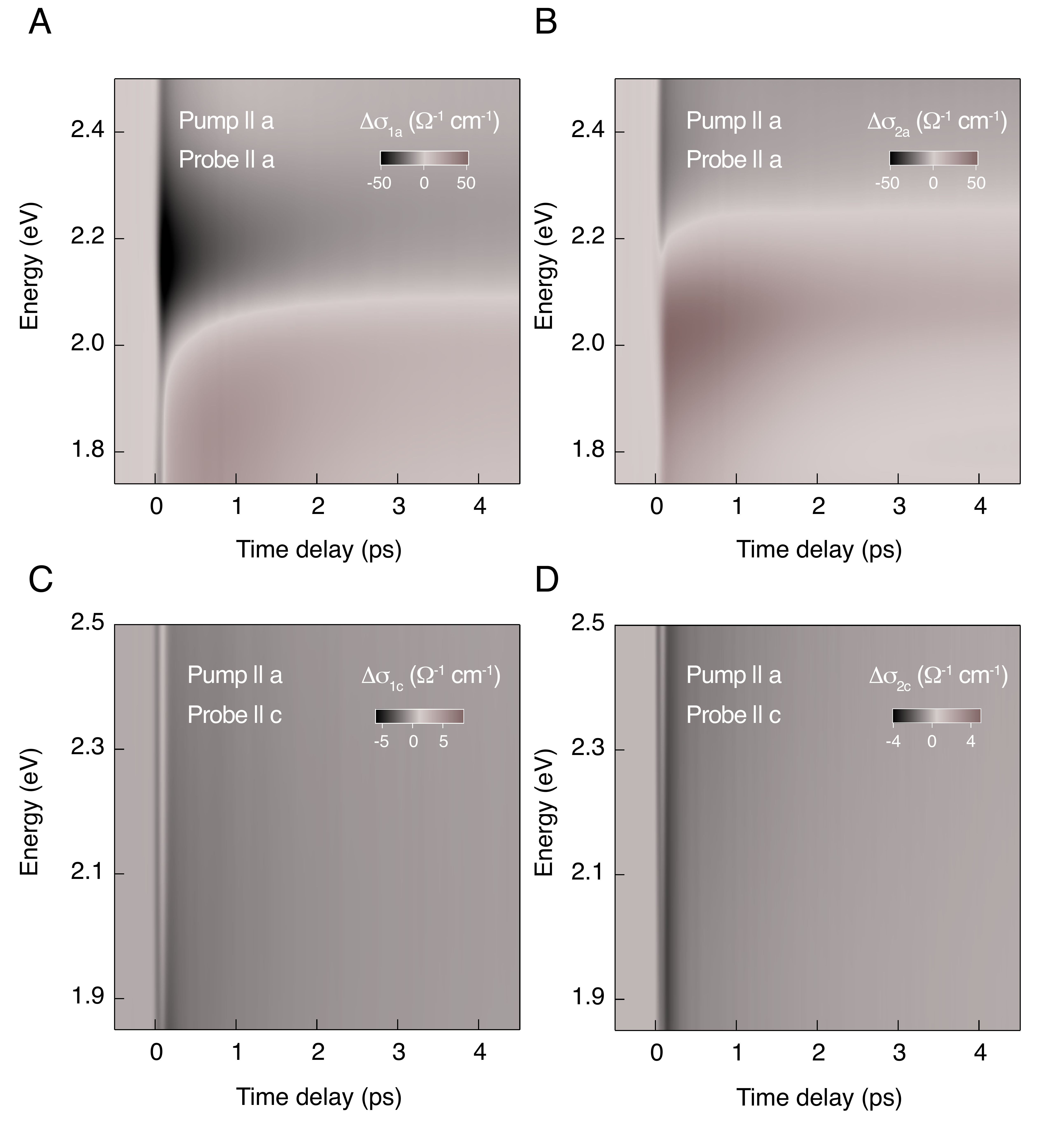}
	\caption{Color-coded maps of the real ($\Delta\sigma_1$) and imaginary ($\Delta\sigma_2$) part of the complex optical conductivity at 10 K with $a$-axis pump polarization and \textit{(A,B)}, $a$-axis, \textit{(C,D)}, $c$-axis probe polarization. The pump photon energy is 3.10 eV and the excitation density is 0.06 photons/Cu.}
	\label{DeltaSigma}
\end{figure*}
\newpage
\clearpage

\begin{figure*}[ht!] 
	\centering
	\includegraphics[width=\columnwidth]{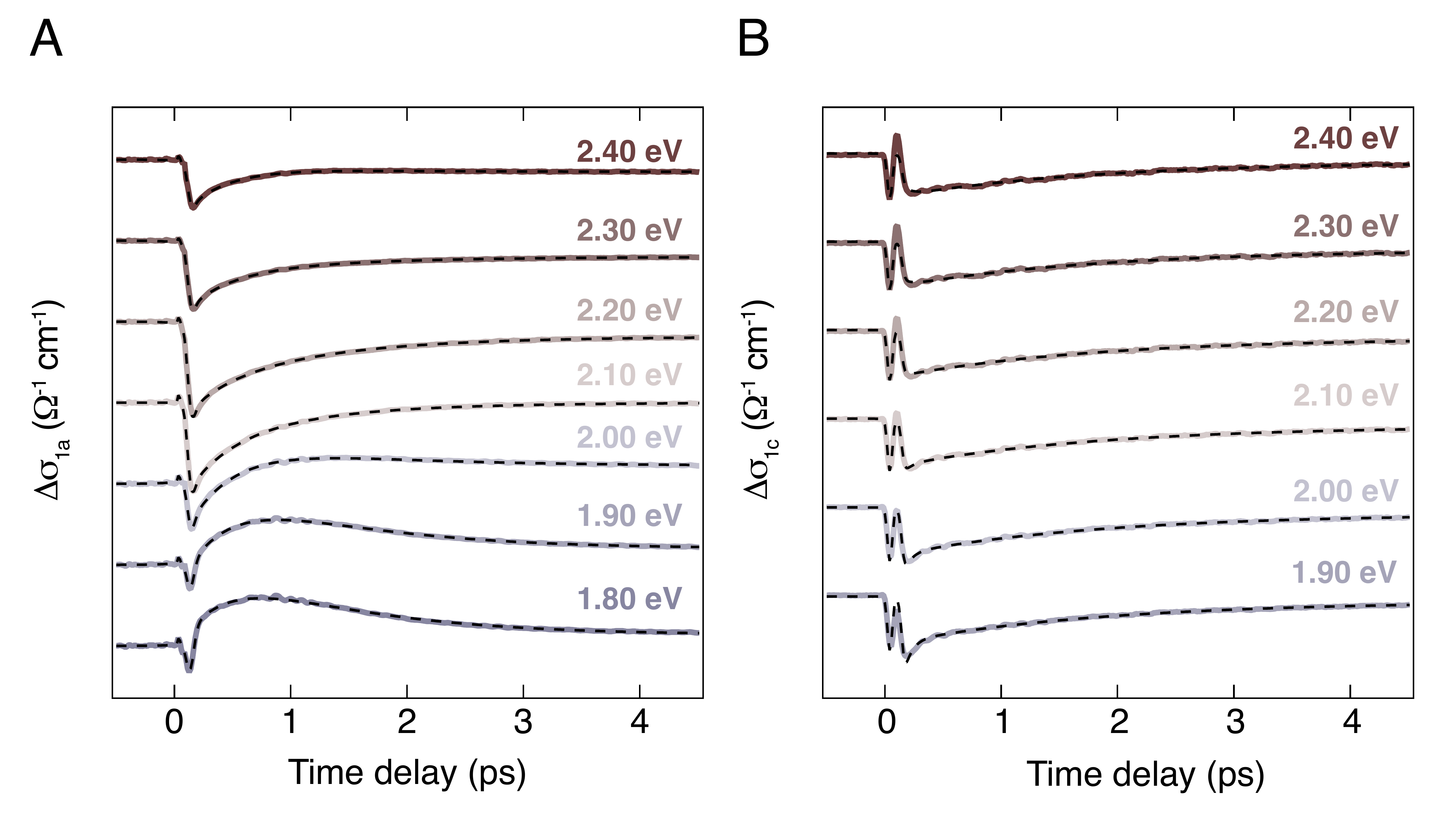}
	\caption{Global fit analysis of the $\Delta\sigma_1$ temporal traces along the \textit{(A)}, $a$-axis, and \textit{(B)}, $c$-axis. The solid colored curves represent the measured response resulting from the integration over 0.10 eV around the indicated photon energy. The dashed black curves on top are the results of the global fit analysis. The curves have been vertically shifted from each other for clarity.}
	\label{GlobalFit}
\end{figure*}
\newpage
\clearpage

\begin{figure*}[ht!] 
	\centering
	\includegraphics[width=\columnwidth]{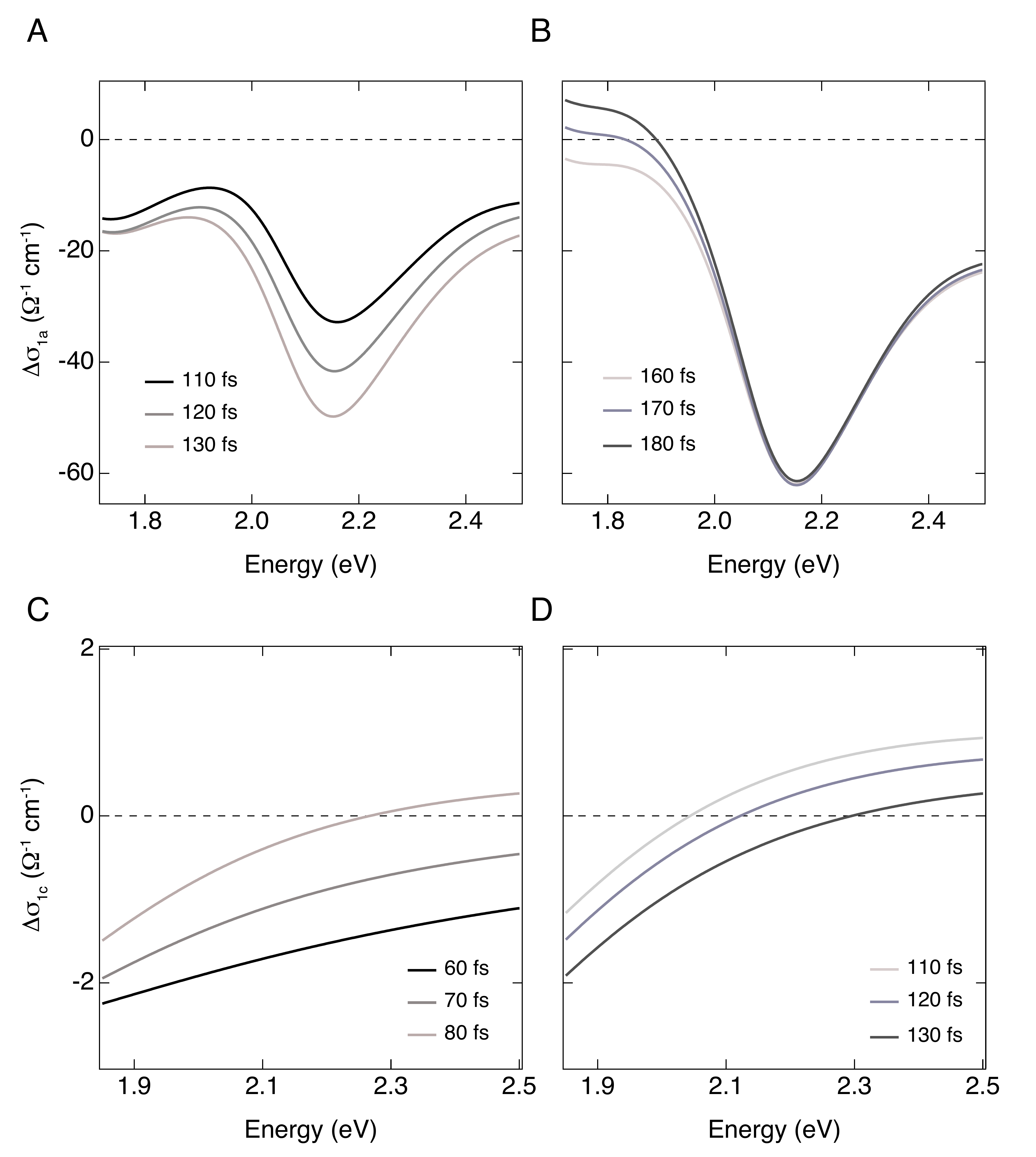}
	\caption{Temporal evolution of the real part of the optical conductivity ($\Delta\sigma_1$) at different pump-probe time delays. The time delays are selected to highlight similarities between the two responses. \textit{(A,B)} In-plane response during the rise of the signal: $(A)$ 110-130 fs; $(B)$ 160–180 fs. \textit{(C,D)} Out-of-plane response during the rise of the signal: $(C)$ 60-80 fs; $(D)$ 110-130 fs.}
	\label{Ds1_spectra}
\end{figure*}
\newpage
\clearpage

\begin{figure*}[ht!] 
	\centering
	\includegraphics[width=\columnwidth]{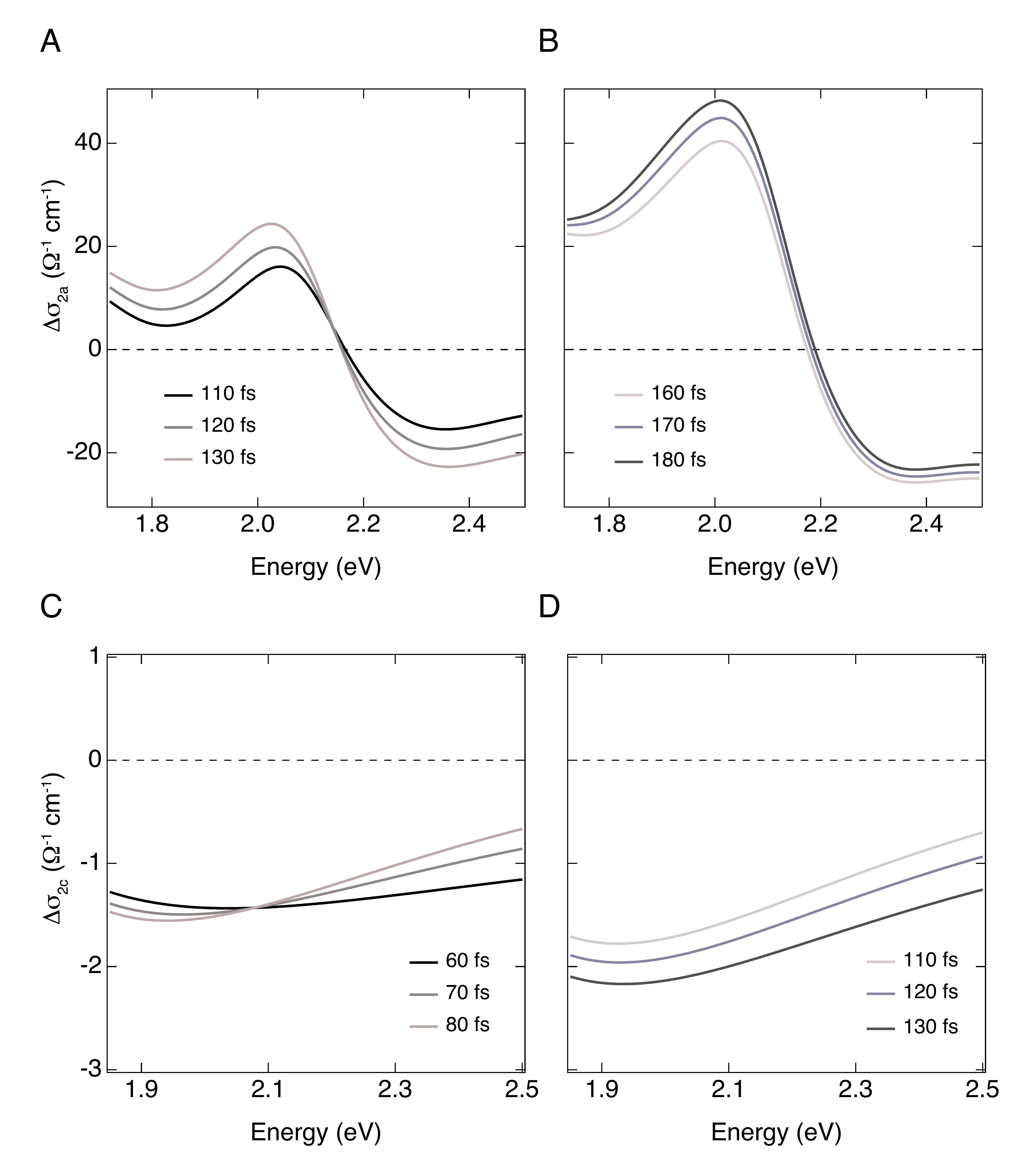}
	\caption{Temporal evolution of the imaginary part of the optical conductivity ($\Delta\sigma_2$) at different pump-probe time delays. The time delays are selected to highlight similarities between the two responses. \textit{(A,B)} In-plane response during the rise of the signal: $(A)$ 110-130 fs; $(B)$ 160–180 fs. \textit{(C,D)} Out-of-plane response during the rise of the signal: (c) 60-80 fs; (d) 110-130 fs.}
	\label{Ds2_spectra}
\end{figure*}
\newpage
\clearpage

\begin{figure*}[ht!] 
	\centering
	\includegraphics[width=0.7\columnwidth]{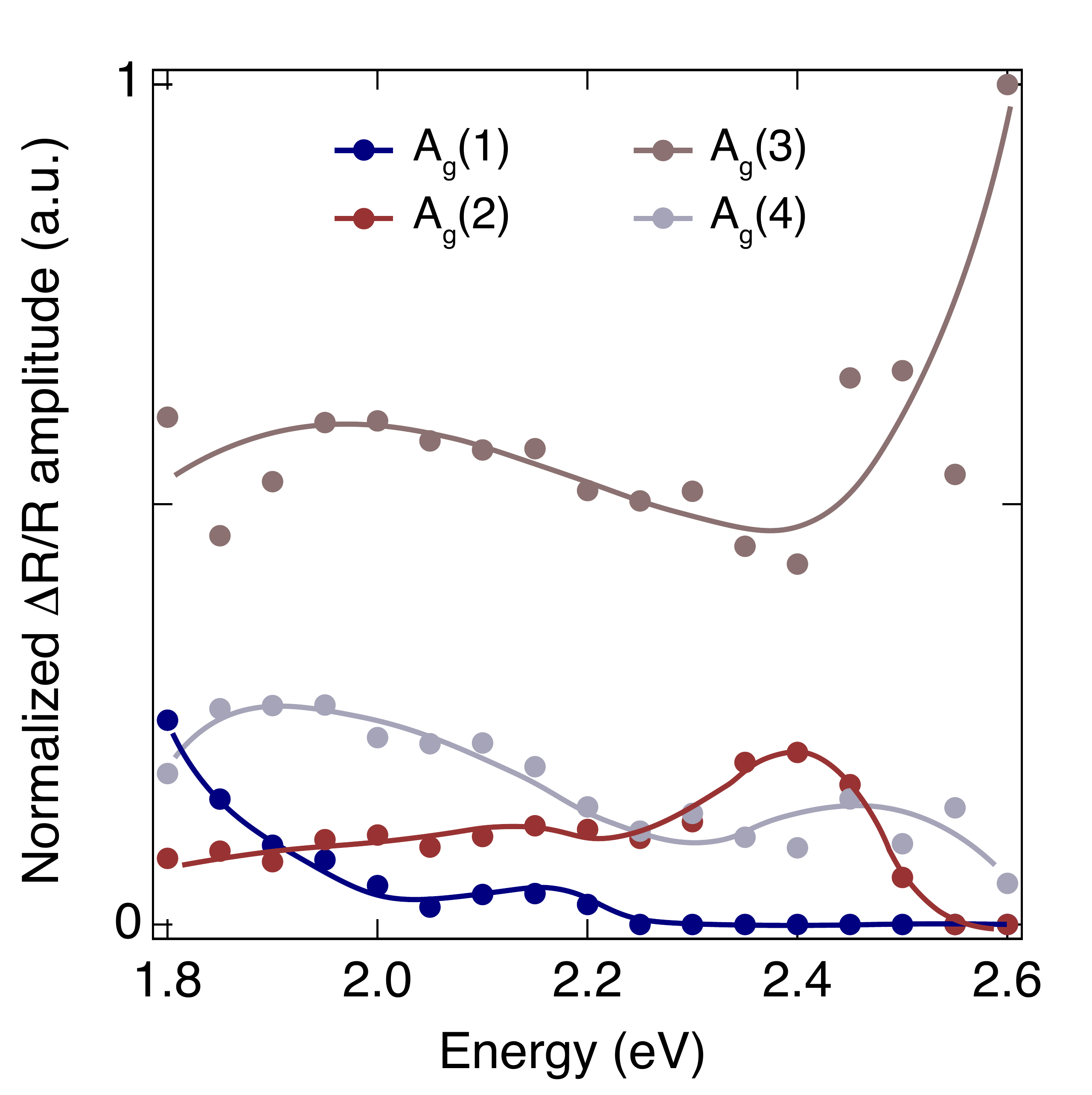}
	\caption{Normalized $\Delta$R/R oscillation amplitude of four distinct $A_g$ Raman-active modes as a function of photon energy when the pump polarization is set along the $a$-axis and the probe polarization along the $c$-axis. Solid lines are used as guides to the eye.}
	\label{Raman}
\end{figure*}
\newpage
\clearpage

\begin{figure*}[ht!] 
	\centering
	\includegraphics[width=0.9\columnwidth]{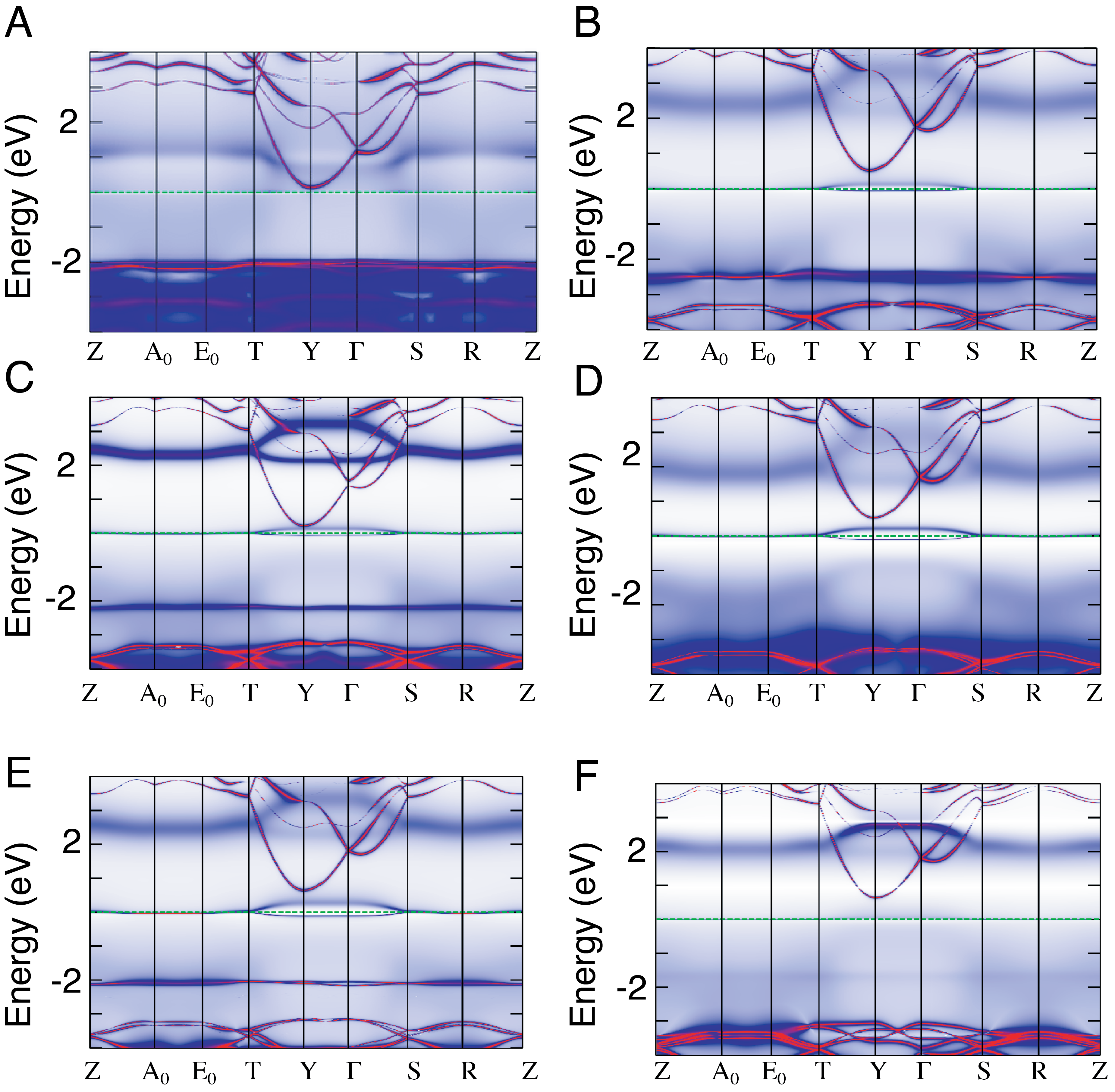}
	\caption{Electronic spectral function $A(k,\omega)$ of the displaced compound obtained in QS$GW$+DMFT and resolved in momentum along a  path in the Brillouin zone connecting the high-symmetry points. The spectral function is obtained for the structures distorted along the eigenvectors of the A$_{g}$ and B$_{g}$ phonon modes calculated by density-functional theory. In particular, the displacements are along \textit{(A)}, the A$_g$(1) mode, \textit{(B)}, the A$_g$(2) mode, \textit{(C)}, the A$_g$(3) mode, \textit{(D)}, the A$_g$(4) mode, \textit{(E)}, the A$_g$(5) mode, and \textit{(F)}, the B$_{3g}$ mode coordinates. The electronic structure in the case of all A$_{g}$ modes exhibits a broad, incoherent 3-peak correlated metallic spectral function, whereas the electronic structure in the case of the B$_{g}$ modes is gapped. The metallicity in the A$_g$(1)-displaced structure is carried by the incoherent spectral weight crossing the Fermi level, putting this displaced structure on the verge of an IMT. In the three other displaced structures, we observe the formation of a weakly-coherent quasiparticle feature around Fermi energy, along with atomic-like Hubbard band features at high energy. The unoccupied coherent band at Y, near the Fermi level, is the La-$d$ orbital band.}
	\label{Quasiparticle}
\end{figure*}
\newpage
\clearpage
\thispagestyle{empty}

\begin{figure*}[ht!] 
	\centering
	\includegraphics[width=\columnwidth]{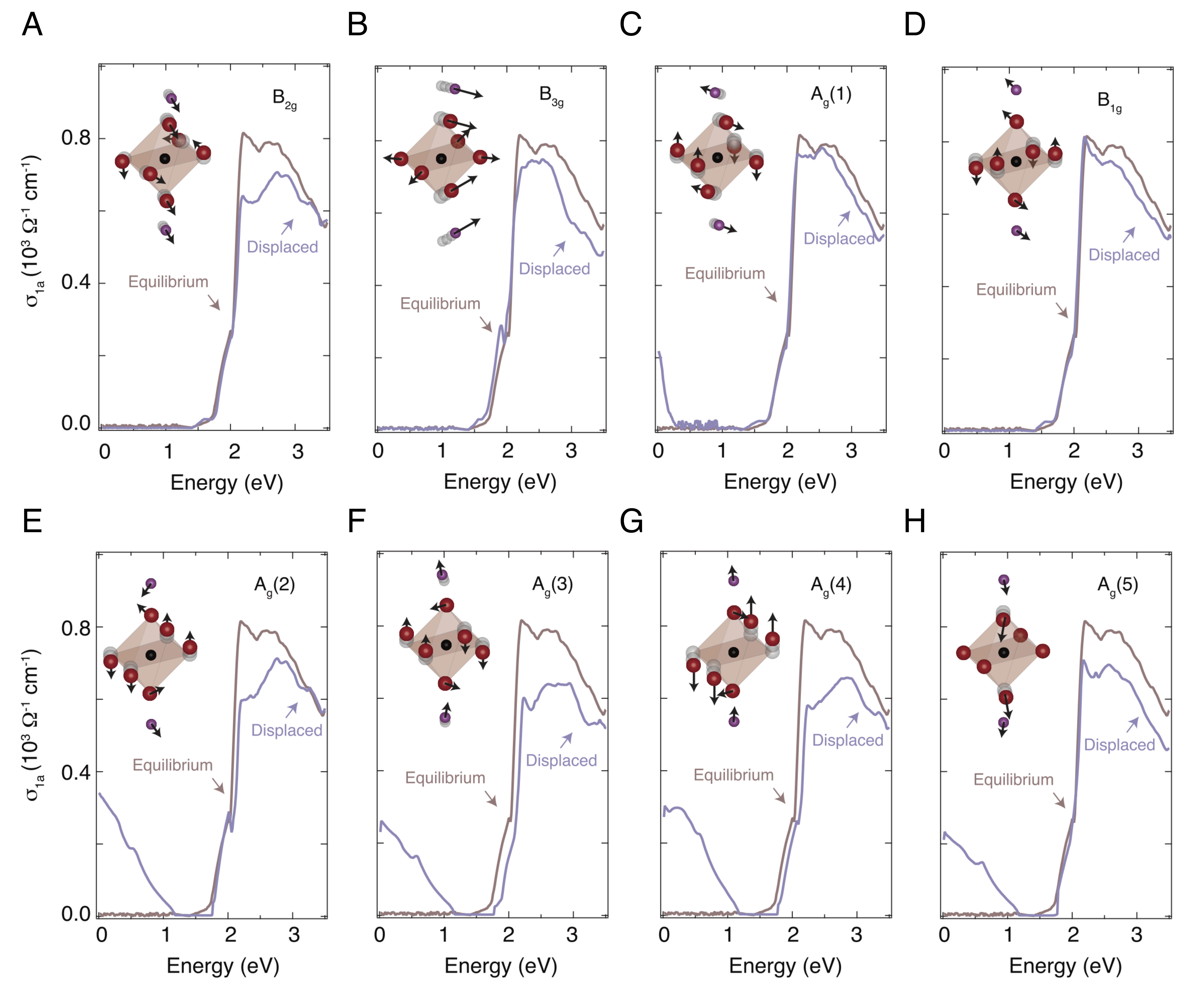}
	\caption{Theory data from QS$GW$+DMFT simulations of in-plane optical conductivity for the LCO unit cell displaced by 0.04 $\AA$ along different phonon coordinates (indicated in each panel). For displacements along totally-symmetric $A_g$ modes, metallization occurs with spectral weight appearing at low energy below $\sim$1.00 eV. In contrast, for displacements along $B_g$ modes, there is no metallization and hardly any impact on the low-energy spectral weight inside the optical CT gap. The total spectral weight is conserved over all energies. For the $B_g$ modes, we find that the missing spectral weight from the optical CT build up shifts towards higher energies.}
	\label{Modes}
\end{figure*}
\newpage
\clearpage

\begin{figure*}[t]
	\centering
	\includegraphics[width=\columnwidth]{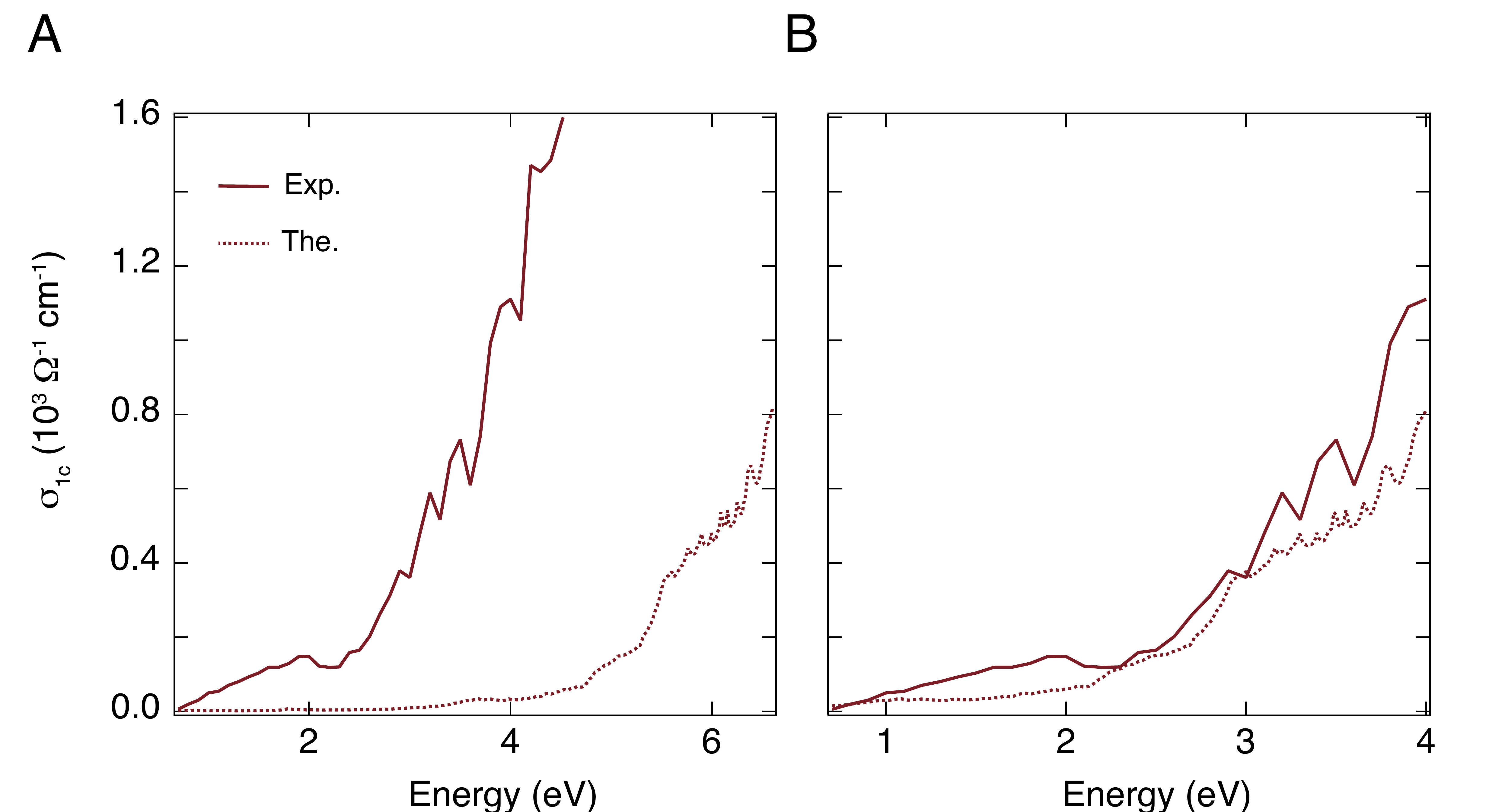}
	\caption{Real part of the optical conductivity at 10 K, measured with the electric field polarized along the $c$-axis (brown solid curve). The QS$GW$+DMFT simulations for the out-of-plane response are shown as a brown dashed curve. (\textit{A}) Comparison between the experimental data and the as-calculated spectrum. The theoretical response is blueshifted with respect to the experimental data due to the lack of electron-hole screening vertex corrections. (\textit{B}) Same data as panel A, but with the theoretical data rigidly redshifted by 2.6 eV. Applying a rigid shift allows to capture the experimental response. We expect the future inclusion of electron-hole scattering vertex correction to provide a contribution similar to this rigid shift.}\label{fig:c_axis}
\end{figure*}
\newpage
\clearpage

\begin{figure*}[ht!] 
	\centering
	\includegraphics[width=\columnwidth]{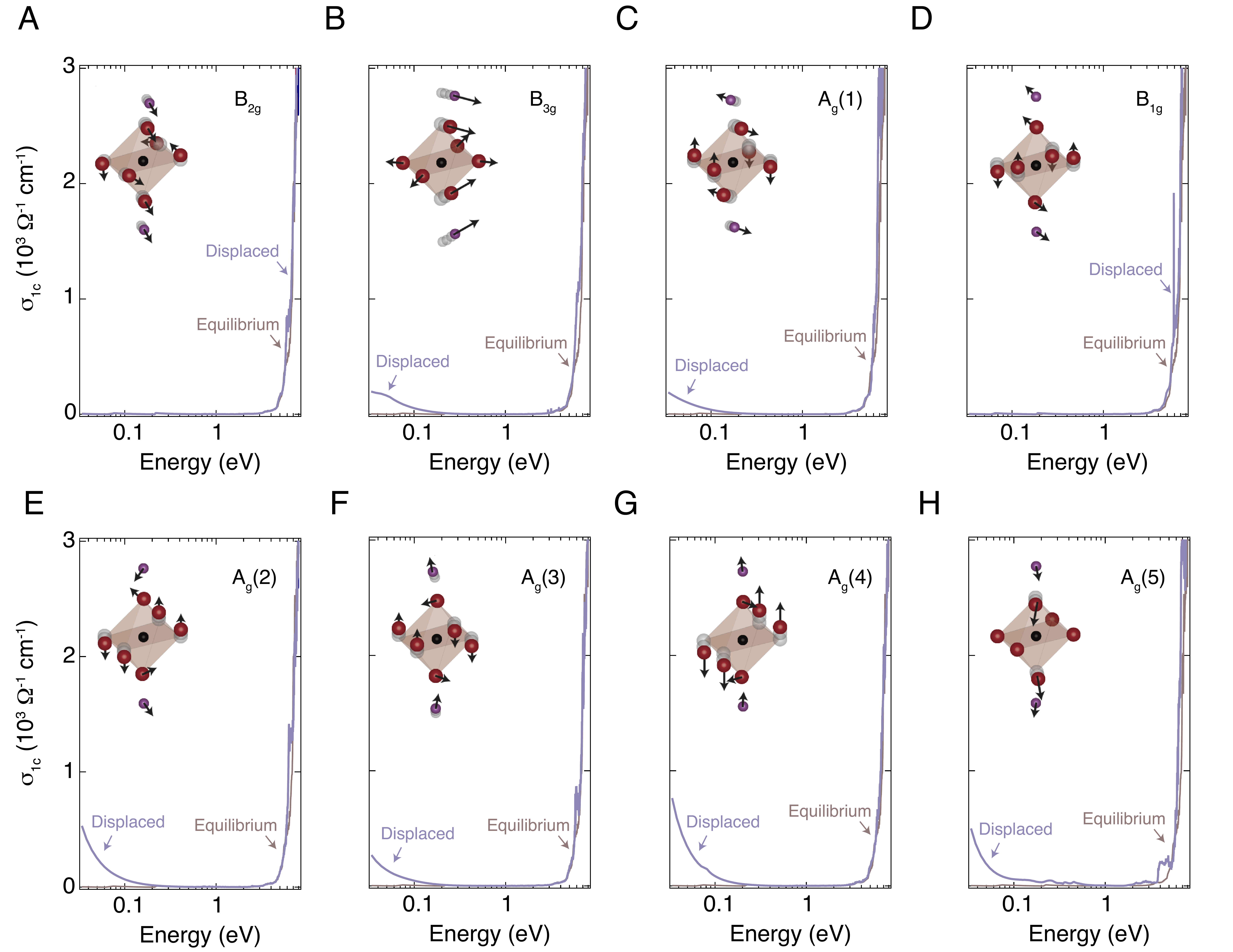}
	\caption{Theory data from QS$GW$+DMFT simulations of out-of-plane optical conductivity for the LCO unit cell displaced by 0.04 $\AA$ along different phonon coordinates (indicated in each panel). For displacements along totally-symmetric $A_g$ modes, metallization occurs with spectral weight appearing at low energy below $\sim$1.00 eV. In contrast, for displacements along $B_{1g}$ and $B_{2g}$ modes, there is no metallization and hardly any impact on the low-energy spectral weight inside the optical CT gap. Finally, displacements along the $B_{3g}$ mode coordinates produces incoherent metallization. The total spectral weight is conserved over all energies. To reveal all the features in the spectrum, a logarithmic scale has been used for the energy axis.}
	\label{Modes_c_axis}
\end{figure*}
\newpage
\clearpage
\newpage


\providecommand{\noopsort}[1]{}\providecommand{\singleletter}[1]{#1}

\end{document}